 \documentclass[aps,pra,preprint,groupedaddress,showpacs]{revtex4}
\usepackage{graphicx}
\usepackage{bm}
\usepackage{floatflt}

\begin{document}
\title{Universal low-energy properties of three two-dimensional particles}


\author{O.~I.~Kartavtsev}
\author{A.~V.~Malykh}
\affiliation{
Joint Institute for Nuclear Research, Dubna, 141980, Russia }


\begin{abstract}

Universal low-energy properties are studied for three identical bosons 
confined in two dimensions. 
The short-range pair-wise interaction in the low-energy limit is described 
by means of the boundary condition model. 
The wave function is expanded in a set of eigenfunctions on the hypersphere 
and the system of hyper-radial equations is used to obtain analytical and 
numerical results. 
Within the framework of this method, exact analytical expressions are derived 
for the eigenpotentials and the coupling terms of hyper-radial equations. 
The derivation of the coupling terms is generally applicable to a variety of 
three-body problems provided the interaction is described by the boundary 
condition model. 
The asymptotic form of the total wave function at a small and a large 
hyper-radius $\rho $ is studied and the universal logarithmic dependence 
$\sim \ln^3\! \rho $ in the vicinity of the triple-collision point is derived. 
Precise three-body binding energies and the $2 + 1$ scattering length 
are calculated. 

\end{abstract}

\maketitle

\section{Introduction}

Dynamics of few particles confined in two dimensions (2D) is of interest 
in connection with numerous investigations ranging from ultra-cold 
gases~\cite{Gorlitz01,Rychtarik04,Petrov00,Bao02} to atoms absorbed 
on a surface~\cite{Hess84,Safonov98,Jarvinen05}. 
An additional motivation is roused by specific features of quantum systems 
in 2D~\cite{Bruch79,Lim80,Verhaar84}. 
The experiments with ultra-cold gases in the 2D and quasi-2D traps have been 
recently realized~\cite{Gorlitz01,Rychtarik04}. 

The description of elementary processes in ultra-cold gases has been 
attracting great interest in the last years and many aspects of the low-energy 
few-body dynamics in three dimensions (3D) have been thoroughly investigated. 
As a particular important example, one could mention the studies of 
the three-body recombination for spinless 
bosons~\cite{Fedichev96,Nielsen99a,Esry99,Macek06}, two-component 
fermions~\cite{Petrov03}, and particles with internal degrees of freedom 
in the presence of a Feshbach resonance~\cite{Kartavtsev02,Petrov04}. 
Low-energy few-body dynamics in low dimensions is less investigated despite  
the extensive studies. 
Besides, new phenomena and additional complications arise in 
quasi-low-dimensional geometry if the effect of motion in the transverse 
directions is taken into account 
(see, e.~g.,~\cite{Petrov00,Wouters03,Idziaszek05,Mora05,Yurovsky06}). 
Among different aspects of 2D systems, one should mention treatment of 
the three-body energy spectra~\cite{Cabral79,Nielsen99,Vranjes02}, 
low-energy scattering of an atom off a dimer molecule~\cite{Adhikari93}, and 
low-energy three-to-three scattering~\cite{Adhikari92,Klemm90}. 
The precise binding energy of four bosons was calculated in~\cite{Platter04} 
and the universal law for the $N$-boson ground-state energy was discussed 
in~\cite{Hammer04}. 

Concerning other 2D and quasi-2D problems, considerable efforts have long been 
devoted to investigation of atoms adsorbed on a surface, 
including helium atoms on graphite~\cite{Dash78} and hydrogen atoms on 
a helium film~\cite{Hess84,Safonov98,Jarvinen05}. 
In this respect, one should mention observation of 
a quasi-condensate~\cite{Safonov98}, measurement of the three-body 
recombination rate~\cite{Jarvinen05} and a vast number of theoretical papers, 
e.~g.,~\cite{Kagan82,Verhaar84,Goey88}.  

In the low-energy limit, which is of interest both for practical applications 
and from the general point of view, the description of the few-body system 
becomes universal, i.~e., essentially independent of the details of 
the two-body interactions. 
Among different results of this sort, notice the recent analytical derivation 
of the universal constants for the zero-energy three-boson scattering in 
3D~\cite{Macek05,Macek06}. 
The description becomes parameterless if only the parameter describing 
the two-body interactions, e.~g., the two-body scattering length $a$, is 
chosen as a scale~\cite{Fedorov93,Kartavtsev99,Braaten03}. 
In comparison with the universal description in 3D, it is of importance that 
the solutions of the three-body problem in 2D , contrary to the 3D one, remain 
regular near the triple-collision point even in the zero-range-interaction 
limit. 
This regularity implies, {\em inter alia}, absence of the Thomas and Efimov 
effect, which was noticed in~\cite{Bruch79,Lim80}. 
Thus, there is no additional regularization parameter, which was introduced in 
3D, and the three-body properties in 2D are completely determined by 
the two-body input that provides a completely universal and parameterless 
description in the low-energy limit. 
In particular, the trimer binding energies and the $2 + 1$ scattering length 
become the universal constants, which must be determined with a good accuracy.

The universal limit corresponds to the limit of the vanishing interaction 
range $r_0$ so that $r_0$ must be much smaller than any length scale in 
the system, i.~e., the binding energies do not exceed the characteristic 
energy $\displaystyle\frac{\hbar^2}{mr_0^2}$. 
Practically, the universal limit can be realized by adjusting the parameters 
of interaction to diminish the two-body binding energy, e.~g., by tuning 
the position of the Feshbach resonance. 
To approach the universal limit, one could use the dependence on the particle 
mass, e.~g., an interesting way is to study the isotopic effect for 2D helium 
atoms~\cite{Cabral79}. 
Generally, the universal limit in 2D appears for a very weak potential with 
a non-positive average as it is known that in this case the two-body binding 
energy in 2D becomes exponentially small with decreasing potential 
strength~\cite{Simon76}. 
It is worthwhile to mention a qualitative difference of the universal 
properties in 2D and 3D, namely, one expects that the both three-body 
bound states arise simultaneously with the two-body bound state, i.~e., at 
the infinite two-body scattering length. 
This conjecture is supported by the calculations~\cite{Cabral79,Blume05}, 
which considered the dependence of the three-body binding energy on 
the particle mass and potential strength in the limit of the vanishing 
two-body binding energy. 
Recall that in 3D an infinite number of three-body bound states arise with 
increasing potential strength at finite values of the scattering length 
before the two-body bound state arises. 

In the present paper, universal properties of three identical 2D bosons 
are studied within the framework of the method which makes use of 
the boundary condition model (BCM)~\cite{Lim80,Wodkiewicz91} for the s-wave 
inter-particle interactions. 
The wave function is expanded in a set of eigenfunctions on 
the hypersphere and a resulting system of hyper-radial equations (HREs) is 
used to conveniently treat both the boundary and scattering problem. 
One of the principal advantages is that for the two-body interaction given by 
the BCM the eigenpotentials of HREs are solutions of a simple eigenvalue 
equation.
In addition, the aim of the present paper is to derive all the terms of HREs 
in the analytical form, thus determining the coupling terms via 
the eigenpotentials and their derivatives over the hyper-radius. 
Both the derivation and the expressions found are generally applicable to 
a variety of problems; as a matter of fact, one can obtain the coupling terms 
in the analytical form for three particles of arbitrary permutation 
symmetry, with arbitrary masses and scattering lengths, and independently of 
the configuration-space dimension. 
In particular, analytical expressions for the coupling terms of the same 
form as in the present paper were derived for three identical bosons in three 
dimensions~\cite{Kartavtsev99}. 
The exact expressions are used to analyze the asymptotic behaviour 
of the coefficients of HREs, their solutions, and the total wave function both 
for large and small inter-particle separation. 
As an important example, the universal dependence of the total wave function 
in the vicinity of the triple-collision point is found. 
In addition, the explicitly known dependence on the channel number is helpful 
to study the role of the channel coupling and to estimate convergence 
of the results with increasing number of HREs. 

Until now, the numerical calculations of the universal constants have 
included the early calculation~\cite{Bruch79} of the binding energies of 
three 2D bosons by solving the momentum-space integral equations. 
Much better precision was obtained by solution of the hyper-radial 
equations~\cite{Nielsen97,Nielsen99}, which results were not in complete 
agreement with those of~\cite{Bruch79}. 
Up to date, in the universal limit of zero-range interactions the most precise 
binding energies have been found by solving the momentum-space integral 
equations~\cite{Hammer04}. 
Among scarce studies of the low-energy three-body scattering in 2D, 
the only available calculation in Ref.~\cite{Adhikari93} demonstrated smooth 
dependence of the $2 + 1$ scattering length on the interaction range. 
In the present paper, the precise universal values of the three-body binding 
energies and the scattering length for a particle collision off a bound pair 
are calculated. 

It has been known for a long time (see, e.~g.,~\cite{McGuire64,Lieb63}) that 
the one-dimensional problem of N identical particles with the zero-range 
interactions (in this case, Dirac's $\delta$-function) is exactly soluble. 
On the other hand, the method of the present paper, including the derivation 
of the exact expressions for the coefficients of HREs, is equally applicable 
to the three-body problem in 1D. 
Note that the approach based on solution of HREs was used in 
Ref.~\cite{Amaya-Tapia98} to discuss low-energy $2 + 1$ scattering in 1D.  
The calculation of the 1D three-body problem provides a good opportunity to 
test the approach, to check the numerical procedure, and to compare the 1D and 
2D calculations. 
For these reasons, the main discussion of the 2D three-body problem is 
complemented by a brief treatment of the corresponding 1D problem. 

The paper is organized in the following manner. 
The next section contains the information on the low-energy two-body 
scattering in 2D, introduces the boundary condition model, and describes 
the expansion of the wave function in a set of eigenfunctions on 
the hypersphere. 
Analytical results are collected in Section~\ref{Analytical}, starting with 
the eigenvalue equation which determines the eigenpotentials for HREs. 
Furthermore, analytical expressions for the coupling terms in HREs are 
derived. 
On the ground of these results, the asymptotic form of the eigenpotentials 
and coupling terms is obtained and applied to derive implications on 
the asymptotic behaviour of the solution of HREs. 
The numerical procedure and the results of the numerical calculations are 
described in Section~\ref{Numerical}, and the last section contains a summary 
and a final conclusion.
The results for three identical bosons in 1D are briefly discussed 
in Appendix.

\section{Method}
\label{Method}

The present study is aimed at the description of the low-energy properties 
of three identical 2D bosons with the short-range pair-wise interaction in 
the limit of the zero interaction range. 
The description turns out to be universal, i.~e., essentially independent of 
the details of the two-body interaction. 
In the low-energy limit under consideration, only the zero total angular 
momentum $L = 0$ should be considered and only the $s$-wave two-body 
interaction should be taken into account. 
The two-body input for the three-body problem is set as the universal 
low-energy description of the two-body interaction by a single parameter, 
for which the two-body scattering length $a$ can be suitably chosen. 
The scattering length in 2D is defined by the asymptotic form of 
the zero-energy wave function at large inter-particle separation $r$ beyond 
the interaction range, $\Psi \sim \ln \frac{r}{a} $~\cite{Lim80,Verhaar84}. 
This is in analogy with the definition of the scattering length in 3D as 
the distance at which the asymptotic expression of the wave function crosses 
zero. 
The $s$-wave scattering amplitude, in accord with the effective-range 
expansion~\cite{Bolle84,Verhaar84,Verhaar85,Adhikari93}, in the low-energy 
limit $k \to 0$ is completely determined by the 2D scattering length $a$, 
\begin{equation}
\label{f0}
f_0(k) = \frac{\sqrt{{2 i}/{\pi k}}}{\cot\delta_0(k) - i} 
\approx \sqrt{\frac{\pi i}{2k}} 
\left[\ln\frac{k a}{2} + \gamma - i\frac{\pi}{2}\right]^{-1} \ , 
\end{equation}
where $k$ is the wave-number, $\delta_0(k) $ is the s-wave scattering phase
shift, and $\gamma \approx 0.5772$ is the Euler constant.

\subsection{Boundary condition model}

In the low-energy limit under consideration, a convenient one-parameter 
description of the two-body interactions is obtained within the framework of 
the BCM if the interaction range is allowed to shrink to zero. 
The two-body interaction introduced in this way is known as the zero-range 
potential~\cite{Demkov88} and the Fermi pseudo-potential~\cite{Wodkiewicz91}. 
The equivalent approach is also used in the momentum-space 
representation~\cite{Braaten03,Hammer04}. 
Within the framework of the BCM corresponding to the vanishing interaction 
range, the exact scattering amplitude $f_0(k)$ is determined by the low-energy 
expression~(\ref{f0})for an arbitrary $k$ and the two-body binding energy 
equals $\displaystyle\frac{4\hbar^2}{ma^2} e^{-2\gamma}$, which corresponds to 
the pure imaginary pole of $f_0(k)$ at $k a = i 2 e^{-\gamma } $. 
Explicitly, the s-wave boundary condition which provides the above-discussed 
low-energy behaviour can be written~\cite{Lim80} as 
\begin{eqnarray}
\label{bound1}
\lim_{r\to 0} \left[ \frac{d}{d r} - \frac{1}{r \ln (r/a)}\right]\Psi = 0\ .
\end{eqnarray}

The total interaction of three particles is a sum of two-body potentials,
which are replaced in the BCM by the two-body boundary condition
of the form~(\ref{bound1}) for each pair of particles.
As the only parameter describing the two-body interactions is the scattering
length $a$, the units $\hbar = m = a = 1$ will be used throughout the paper,
thereby the three-body problem becomes parameterless.
The total wave function $\Psi$ satisfies the boundary conditions and
the Helmholtz equation,
\begin{equation}
\label{shred}
\left[\Delta_{{\mathbf x}} + \Delta_{{\mathbf y}}+E\right]\Psi = 0 \ ,
\end{equation}
where ${\mathbf x}$, ${\mathbf y}$ is an arbitrary pair of the scaled Jacobi
coordinates defined via the particles' radius-vectors ${\mathbf r}_i$ as
${\mathbf x}_i = {\mathbf r}_j - {\mathbf r}_k $ and ${\mathbf y}_i =
\displaystyle\frac{1}{\sqrt{3}}\left(2{\mathbf r}_i - {\mathbf r}_j -
{\mathbf r}_k\right)$.
Different sets of the Jacobi coordinates are related by ${\mathbf x}_i =
-c {\mathbf x}_j + s {\mathbf y}_j $ and ${\mathbf y}_i = -s{\mathbf x}_j -
c{\mathbf y}_j $, where $c = 1/2$, $s = \pm\sqrt{3}/2$, and the $\pm$ sign
is chosen if $ \{ ijk \} $ is an even or odd permutation of $ \{ 123 \} $.
The wave function $\Psi$ of three identical particles is symmetrical
under any permutation of the particles, therefore, it is sufficient to impose 
just one boundary condition,
\begin{eqnarray}
\label{bound2}
\lim_{x \to 0}{\left[\frac{\partial}{\partial x} -
\frac{1}{x \ln x}\right]\Psi} = 0 \ ,
\end{eqnarray}
where $x$ is any of three inter-particle distances.

\subsection{Hyper-radial expansion}
\label{hrexpansion}

Solution of a system of HREs provides an efficient approach to treat both 
the eigenvalue and scattering problem for the three-body 
system~\cite{Nielsen99,Kartavtsev99}. 
This approach is particularly advantageous due to the use of the BCM since all 
the terms of HREs are expressed in the analytical form, which allows one to 
obtain the exact asymptotic form of the wave function and to improve 
the accuracy of the numerical calculations. 
The system of HREs is obtained by expanding the total wave function in a set 
of eigenfunctions on the hypersphere $\Phi_n(\alpha , \theta , R)$,
\begin{equation}
\label{Psi}
\Psi = e^{-R} \sum_{n=1}^{\infty} f_n(R)\Phi_n(\alpha , \theta , R) \ , 
\end{equation}
where the hyper-spherical variables $\rho$ ($0 \leq \rho < \infty$), 
$\alpha_i$ ($0 < \alpha_i \leq \pi/2$), and $\theta_i$ 
($0 < \theta_i \leq \pi$) are introduced by the relations 
$x_i = \rho\sin \alpha_i$, $y_i = \rho\cos \alpha_i$, and
$\cos\theta_i = ({\mathbf x}_i{\mathbf y}_i)/{x_i y_i}$ and $R = \ln\rho$ is 
a convenient variable in 2D. 
Different sets of the hyper-spherical variables are related by 
$\cos{2\alpha_i} = -c \cos{2\alpha_j} + s \sin{2\alpha_j}\cos{\theta_j}$ and 
$\sin 2\alpha_i\cos \theta_i = \pm s \cos 2\alpha_j - 
c \sin 2\alpha_j \cos\theta_j$. 
By definition, $\Phi_n(\alpha , \theta , R)$ are regular solutions of 
the eigenvalue problem on the hypersphere, i.~e., at fixed $R$, deduced from 
Eqs.~(\ref{shred}) and (\ref{bound2}) 
\begin{eqnarray}
\label{eqonhypershere}
&&\left[\Lambda^2 + \xi_n^2(R) - 1\right]\Phi_n(\alpha, \theta, R) = 0 \ , \\ 
\label{bch}
&&\lim_{\alpha \to 0}{\left[\frac{\partial}{\partial \alpha } - 
\frac{1}{\alpha (R + \ln\alpha )}\right]\Phi_n(\alpha ,\theta , R)} = 0 \ , 
\end{eqnarray}
where
\begin{equation}
\label{lambda2}
\Lambda^2 = \frac{\partial^2}{\partial \alpha^2} + 2\cot 
2\alpha\frac{\partial}{\partial\alpha} + \frac{4}{\sin^22\alpha} 
\frac{\partial^2}{\partial \theta^2} \ . 
\end{equation}
Like the total wave function, the functions $\Phi_n(\alpha, \theta, R)$ are 
symmetrical under any permutation of particles, i.~e., 
$\Phi_n(\alpha, \theta, R)$ are independent of the index enumerating 
the Jacobi coordinates. 

For each value of the variable $R$, the problem (\ref{eqonhypershere}), 
(\ref{bch}) determines an infinite number of discrete eigenvalues $\xi_n^2(R)$ 
and corresponding eigenfunctions $\Phi_n$ normalized by the condition 
$\langle\Phi_n|\Phi_m\rangle = \delta_{nm}$. 
Henceforth the notation $\langle\cdot|\cdot\rangle$ means integration over 
the invariant volume on the hypersphere 
$d\Omega = \frac{1}{12}\sin{2\alpha}\,d\alpha\,d\cos\theta$, where 
the arbitrarily chosen factor $1/12$ is suitable for the derivation of 
the coupling terms in Section~\ref{Derivation}. 
The expansion~(\ref{Psi}) of the total wave function leads to a system of 
HREs which can be written in two equivalent forms, 
\begin{eqnarray}
\label{system1}
&& \left[-\frac{d^2}{d R^2} - {\mathrm Q}(R)\frac{d}{d R} - 
\frac{d}{dR}{\mathrm Q}(R) + {\mathrm U}(R) + {\mathrm P}(R) - 
Ee^{2R}\right]{\mathrm f}(R) = 0 \ ,  \\ 
\label{system2}
&& \left[-\left(\frac{d}{dR} + {\mathrm Q}(R)\right)^2 + {\mathrm U}(R) - 
Ee^{2R}\right]{\mathrm f}(R) = 0\ , 
\end{eqnarray}
where ${\mathrm f}(R)$ is the vector-function composed of the hyper-radial 
channel functions $f_n(R)$ and the matrices of eigenpotentials 
${\mathrm U}(R)$ and coupling terms ${\mathrm Q}(R)$ and ${\mathrm P}(R)$ 
are defined by their matrix elements 
\begin{eqnarray}
\label{Unm0}
&&U_{nm}(R) = \xi_n^2(R)\delta_{nm}, \\ 
\label{Qnm0}
&&Q_{nm}(R) = \langle\Phi_n|\Phi_m'\rangle, \\ 
\label{Pnm0} 
&&P_{nm}(R) = \langle\Phi_n'| \Phi'_m\rangle \ , 
\end{eqnarray}
with the prime denoting the derivative over $R$. 
The identity 
\begin{equation}
\label{Pprop1}
P_{nm} = \sum_{k=1}^{\infty}Q_{nk}Q_{mk} 
\end{equation}
provides the equivalence of the infinite systems of equations in the 
forms~(\ref{system1}) and~(\ref{system2}). 

Although two infinite systems of HREs~(\ref{system1}) and~(\ref{system2}) are 
equivalent, the truncated ones give rise to different results, which allows 
one to estimate convergence with increasing number of HREs $N$ in practical 
calculations. 
Notice that $N$ HREs of the form~(\ref{system1}) reduce to 
the form~(\ref{system2}) if the $N$-dimensional matrix ${\mathrm P}^{(N)}$ is 
replaced by a product of $N$-dimensional matrices ${\mathrm Q}^{(N)}$, 
${\mathrm P}^{(N)} \rightarrow -{\mathrm Q}^{(N)}{\mathrm Q}^{(N)}$. 
It is important that the solution of the truncated system of $N$ HREs taken 
in the form~(\ref{system1}) gives the upper bound $E_i^{(N)}$ for 
the exact energy of the $i$th state $E_i$ and the upper bound $A^{(N)}$ for 
the exact scattering length $A$, i.e., $E_i^{(N)}\geq E_i$ and 
$A^{(N)}\geq A$~\cite{Starace79,Coelho91}. 
The proof can be obtained by observing that the truncated system of HREs 
in the form~(\ref{system1}) can be obtained by application of 
the variational principle with the trial function containing a finite sum of 
the form~(\ref{Psi}). 
On the other hand, the solution of HREs~(\ref{system2}), at least in 
the one-channel approximation, gives the lower bound for the ground-state 
energy~\cite{Starace79}. 
Solution of the system~(\ref{system1}) generally provides faster convergence 
with increasing number of equations, while solution of 
the system~(\ref{system2}) does not require elaborate calculation of 
$P_{nm}(R) $. 
Notice that the scattering length can be calculated by solving only the 
truncated system~(\ref{system1}) because the first-channel effective potential
$U^{eff}_1(\rho) $ decreases as $1/\rho^4$ (see Section~\ref{Asymptotic}).
In contrast to that, the first-channel effective potential in the truncated 
system~(\ref{system2}) is of the form $U^{eff}_1(\rho) = \left[\xi_1^2(\rho) +
\sum_{n=1}^N Q_{1n}^2(\rho)\right]/\rho^2$ and contains a long-range term 
$\sim -1/3\rho^2$ for any finite $N$, which prevents calculation of 
the scattering length. 

\section{Analytical results}
\label{Analytical}

\subsection{Eigenvalue problem on the hypersphere}
\label{Eigenvalue}

It is convenient to take account of the permutation symmetry and to satisfy 
the boundary condition~(\ref{bch}) by means of the Faddeev-like decomposition, 
\begin{eqnarray}
\label{faddeev}
\Phi(\alpha ,\theta , R) = \sum_{i=1}^{3}\chi(\alpha_i, R) \ ,
\end{eqnarray}
provided the function $\chi(\alpha_i, R)$ is symmetrical under the permutation 
of the particles $j$ and $k$ and satisfies the same equation on 
a hypersphere~(\ref{eqonhypershere}) as the eigenfunction 
$\Phi(\alpha ,\theta , R)$. 
The representation~(\ref{faddeev}) is advantageous due to a simple structure 
of the function $\chi(\alpha, R)$, which is singular only at one point 
$\alpha = 0$ and does not depend on $\theta $ because of the $s$-wave boundary 
condition. 
Following Eq.~(\ref{bch}), the boundary condition for the function 
$\chi(\alpha_i, R)$ takes the form, 
\begin{eqnarray}
\label{bound}
\lim_{\alpha_i\to 0}{\left[ \frac{\partial\chi(\alpha_i , R)} 
{\partial \alpha_i} - \frac{1} {\alpha_i(R + \ln\alpha_i)} 
\sum_{j=1}^{3}\chi(\alpha_j , R) \right]} = 0\ , 
\end{eqnarray}
where the sum contains two functions $\chi (\alpha_j, R)$ (for $j \ne i$), 
which are regular in the limit $\alpha_i \to 0$. 
The solution to the eigenvalue problem on the hypersphere is straightforward 
in terms of the Legendre function $P_\nu(x)$ regular at 
$x = 1$~\cite{Bateman53}, 
\begin{eqnarray}
\label{chi}
\chi(\alpha , R) = A(R) P_{\frac{\xi(R) - 1}{2}}\left(-\cos 2\alpha\right) \ ,
\end{eqnarray}
where $A(R)$ is the normalization constant. 
Substituting~(\ref{chi}) into the boundary condition~(\ref{bound}), 
using the asymptotic form of the Legendre function as $\alpha \to 0$, 
$P_\nu(-\cos 2\alpha)\to \displaystyle\frac{2}{\pi}\sin\pi\nu\left[\ln\alpha +
\gamma + \psi(\nu + 1)\right] + \cos\pi\nu$~\cite{Bateman53}, 
and calculating the limit $\cos 2\alpha_{j,k} \to -1/2$ as $\alpha_i \to 0$, 
one comes to the eigenvalue equation, 
\begin{equation}
\label{transeq}
R - \gamma - \psi\left(\frac{\xi+1}{2}\right) +
\frac{\pi}{2}\tan\frac{\pi}{2}\xi +
{\pi}{\sec\frac{\pi}{2}\xi}P_{\frac{\xi-1}{2}}\left(\frac{1}{2}\right) = 0 \ ,
\end{equation}
where $\psi (x)$ is the digamma function. 
The same eigenvalue equation, in slightly different notation, was derived 
in Ref.~\cite{Nielsen97} in the limit of the zero interaction range. 

Considering the solution of Eq.~(\ref{transeq}), it is worthwhile to note that 
the left-hand side is an even function of $\xi$, i.e., $R$ is a function of 
$\xi^2$. 
Similar to the corresponding eigenvalue equation in 3D 
\cite{Fedorov93,Kartavtsev99}, the transcendental equation~(\ref{transeq})
determines the infinitely multivalued function $\xi^2 (R)$ for an arbitrary
complex-valued variable $R$.
In particular, different branches of this unique function for the real-valued
$R$ form a set of the real-valued $\xi_n^2(R)$ which play the role of
eigenpotentials in the HREs.
Hereafter it is convenient to enumerate $\xi_n^2(R)$ by an index
$n = 1, 2, 3,\dots$ in ascending order.
As $R$ increases from $-\infty$ to $\infty$, all the terms $\xi_n^2(R)$
decrease monotonically in the intervals $-\infty < \xi_1^2(R) < 1 $,
$1 < \xi_2^2(R) < 25 $, and $(2n - 1)^2 < \xi_n^2(R) < (2n + 1)^2 $
for $n > 2$.
Note that at the exceptional point $\xi = 3$ the solution of the eigenvalue 
equation~(\ref{transeq}) gives a finite limit $R_0 \approx 1.64$; 
nevertheless, calculation of the function $\xi_2^2(R)$ and its derivatives in 
the vicinity of this point requires a special care to take into account 
exact cancellation of divergent terms. 

\subsection{Derivation of the coupling terms }

\label{Derivation}

An important advantage of the BCM is the analytical expression~(\ref{transeq}) 
for the eigenpotentials $\xi_n^2(R)$ in HREs that allows one to 
study the asymptotic properties of the solution and to simplify the numerical 
calculation thus improving its accuracy. 
Evidently, the analytical expressions are strongly desirable for the coupling 
terms $Q_{nm}(R)$ and $P_{nm}(R)$. 
Whereas the direct evaluation of $Q_{nm}(R)$ and $P_{nm}(R)$ by means of 
the definitions (\ref{Qnm0}),~(\ref{Pnm0}) is quite involved, fortunately, one 
can circumvent this obstacle by using the explicit dependence on the parameter 
$R$ in the eigenvalue problem (\ref{eqonhypershere}), (\ref{bch}). 
Thus, within the framework of the BCM one derives the analytical expression 
for $Q(\rho)$ and $P(\rho)$ via eigenpotentials $\xi_n^2(R)$ and their 
derivatives over $R$. 

To simplify the notation, the eigenvalue problem
on the hypersphere~(\ref{eqonhypershere}), (\ref{bch}) is written as
\begin{eqnarray}
\label{eqonR1}
&&(\Lambda^2 - \varepsilon_n)\Phi_n = 0\ , \\
\label{boundPhi1}
&&\lim_{\alpha_i\to 0}\left(\frac{\partial\Phi_n}{\partial \alpha_i} -
\frac{\phi_n}{\alpha_i}\right) = 0 \ ,
\end{eqnarray}
where $\varepsilon_n = -\xi_n^2 + 1$, and the function 
\begin{eqnarray}
\label{defphi}
\phi_n(\alpha, R) = \frac{\Phi_n(\alpha, R)}{\ln\alpha + R}
\end{eqnarray}
tends, as $\alpha \to 0$, to the finite limit which does not depend on 
the index enumerating the different sets of the Jacobi coordinates. 
Taking the derivatives of Eqs.~(\ref{eqonR1}) and (\ref{boundPhi1}) with 
respect to $R$, one obtains that $\Phi_n'$ satisfy the inhomogeneous equation 
\begin{equation}
\label{eqonR2}
(\Lambda^2 - \varepsilon_n)\Phi_n' = \varepsilon_n'\Phi_n
\end{equation}
and the boundary condition
\begin{equation}
\label{boundPhi2}
\lim_{\alpha_i\to 0}\left(\frac{\partial\Phi_n'}{\partial \alpha_i} -
\frac{\phi_n'}{\alpha_i}\right) = 0 \ .
\end{equation}

For derivation of $Q_{nm}(R)$, one starts with the Hellmann-Feynman-type
relation
\begin{eqnarray}
\label{Qn1}
\langle\Phi_m|\Lambda^2|\Phi'_n\rangle -
\langle\Phi_n'|\Lambda^2|\Phi_m\rangle = \varepsilon'_n \delta_{nm} +
(\varepsilon_m-\varepsilon_n)Q_{nm} \ ,
\end{eqnarray}
which is obtained by projecting Eqs.~(\ref{eqonR1}) and (\ref{eqonR2}) onto 
the functions $\Phi_m'$ and $\Phi_m$, respectively. 
On the other hand, the integrals over the hypersphere on the left-hand side 
of Eq.~(\ref{Qn1}) reduce to the contour integrals around three points 
$\alpha_i = 0 $ in which the functions $\Phi_n$ and $\Phi_n'$ have
singularities. 
Allowing the length of the contours to shrink to zero and taking into account 
that all three singular points $\alpha_i=0$ make equal contributions for 
the symmetry reason, one obtains 
\begin{eqnarray}
\label{Qn1a}
\langle\Phi_m|\Lambda^2|\Phi'_n\rangle -
\langle\Phi_n'|\Lambda^2|\Phi_m\rangle =
\lim_{\alpha\to 0}\alpha\left[\Phi_n' \frac{\partial \Phi_m}{\partial\alpha} -
\Phi_m\frac{\partial \Phi_n'}{\partial\alpha}\right]\ .
\end{eqnarray}
Combining the boundary conditions~(\ref{boundPhi1}),~(\ref{boundPhi2}) with
Eq.~(\ref{defphi}), one finds
\begin{eqnarray}
\label{Qn1b}
\lim_{\alpha\to 0}\alpha\left[\Phi_m\frac{\partial \Phi_n'}{\partial\alpha} -
\Phi_n' \frac{\partial \Phi_m}{\partial\alpha}\right] =
\phi_n(0, R)\phi_m(0, R)
\end{eqnarray}
and eventually comes from (\ref{Qn1})-(\ref{Qn1b}) to the basic relation
\begin{eqnarray}
\label{Qn2}
\varepsilon'_n \delta_{nm} + (\varepsilon_m - \varepsilon_n) Q_{nm} -
\phi_n(0, R)\phi_m(0, R) = 0 \ .
\end{eqnarray}
The diagonal part of (\ref{Qn2}) provides a simple relation between
$\phi_n(0, R)$ and $\varepsilon_n'$,
\begin{eqnarray}
\label{varphi}
\varepsilon'_n - \phi_n^2(0, R) = 0 \ ,
\end{eqnarray}
while the non-diagonal part of (\ref{Qn2}) combined with (\ref{varphi}) gives
finally the desired result
\begin{eqnarray}
\label{Qn3}
Q_{nm} =
\frac{\sqrt{\varepsilon_n'\varepsilon_m'}}{\varepsilon_m - \varepsilon_n} \ .
\end{eqnarray}

In a similar way, to derive $P_{nm}(R)$ for $n\neq m$, one calculates 
the difference $\langle\Phi_m'|\Lambda^2|\Phi'_n\rangle - 
\langle\Phi_n'|\Lambda^2|\Phi_m'\rangle$ by projecting Eq.~(\ref{eqonR2}) 
onto the functions $\Phi'_{n,m}$ and integrating on the hypersphere, which 
gives 
\begin{eqnarray}
\label{Pn1}
(\varepsilon_n - \varepsilon_m)P_{nm} +
(\varepsilon_n' + \varepsilon_m')Q_{nm} =
- \lim_{\alpha \to 0}\alpha\left[\Phi'_m\frac{\partial \Phi_n'}
{\partial\alpha} - \Phi_n' \frac{\partial \Phi_m'}{\partial\alpha}\right] \ .
\end{eqnarray}
In view of (\ref{boundPhi1}) and (\ref{boundPhi2}), the limit on 
the right-hand side of Eq.~(\ref{Pn1}) equals $\phi_m(0, R)\phi_n'(0, R) -
\phi_n(0, R)\phi_m'(0, R)$, which allows one to obtain, by expressing
$\phi_n(0, R)$, $\phi_n'(0, R)$ via $\varepsilon_n'$, $\varepsilon_n''$ from
Eq.~(\ref{varphi}),
\begin{eqnarray}
\label{Pn2}
P_{nm} = Q_{nm}\left[\frac{\varepsilon_n' + \varepsilon_m'}
{\varepsilon_m - \varepsilon_n} + \frac{1}{2}\left(\frac{\varepsilon_n''}
{\varepsilon_n'} - \frac{\varepsilon_m''}{\varepsilon_m'}\right)\right]\ .
\end{eqnarray}

For derivation of the diagonal terms $P_{nn}(R)$, one requires the functions 
$\Phi_n''$, which satisfy the inhomogeneous equation 
\begin{equation}
\label{eqonR3}
(\Lambda^2 - \varepsilon_n)\Phi_n'' = 2 \varepsilon_n'\Phi_n'
+ \varepsilon_n''\Phi_n
\end{equation}
and the boundary condition
\begin{equation}
\label{boundPhi3}
\lim_{\alpha_i\to 0}\left(\frac{\partial\Phi_n''}{\partial\alpha_i } -
\frac{\phi_n''}{\alpha_i}\right) = 0 \ .
\end{equation}
Repeating the above procedure to calculate the difference
$\langle\Phi_n'|\Lambda^2|\Phi_n''\rangle -
\langle\Phi_n''|\Lambda^2|\Phi'_n\rangle $ and taking into account
the identity $P_{nn} = -\langle\Phi_n''|\Phi_n\rangle$ one obtains
\begin{eqnarray}
\label{Pn3}
3\varepsilon'_n P_{nn} = \lim_{\alpha\to 0} \alpha
\left[\Phi_n'\frac{\partial \Phi_n''} {\partial\alpha} -
\Phi_n''\frac{\partial \Phi_n' }{\partial\alpha}\right] =
2[\phi_n'(0, R)]^2 - \phi_n(0, R)\phi_n''(0, R)  \ ,
\end{eqnarray}
which after simple algebra in combination with (\ref{varphi}) gives rise to
\begin{eqnarray}
\label{Pn4}
P_{nn} = -\frac{1}{6}\frac{\varepsilon_n'''}{\varepsilon_n'} + \frac{1}{4}
\left(\frac{\varepsilon_n''}{\varepsilon_n'}\right)^2\ .
\end{eqnarray}

The derivation of all the terms in HREs is accomplished with the exact
expressions (\ref{Qn3}), (\ref{Pn2}), and (\ref{Pn4}) for the coupling terms
$Q_{nm}(R)$ and $P_{nm}(R)$ and the eigenvalue equation~(\ref{transeq})
for $\xi^2(R)$.
Whereas the explicit value of $\phi_n(0, R)$ is of no interest for
determination of $Q_{nm}(R)$ and $P_{nm}(R)$, it is easy to calculate
the limit $\alpha \to 0$ in Eq.~(\ref{defphi})
\begin{eqnarray}
\phi_n(0, R) = \frac{2}{\pi}A_n\cos\frac{\pi}{2}\xi_n \ ,
\end{eqnarray}
which, in view of Eq.~(\ref{varphi}), allows the normalization constant
to be additionally determined,
\begin{eqnarray}
\label{normconst}
A_n = \frac{\pi}{2}\sqrt{- 2 \xi_n \xi_n'} \sec\frac{\pi}{2}\xi_n \ .
\end{eqnarray}

One should emphasize generality of the derived analytical 
expressions~(\ref{Qn3}), (\ref{Pn2}), and (\ref{Pn4}). 
The derivation is based essentially on the BCM used to describe the pair-wise 
interaction. 
Within the framework of the BCM the described procedure is applicable to 
derivation of the coupling terms for a variety of three-body systems in 
the configuration space of an arbitrary dimension including particles of 
different masses and scattering lengths and particles with internal degrees 
of freedom. 
In particular, the analytical expressions of the same form are valid 
for three identical bosons in 3D~\cite{Kartavtsev99} and in 1D (discussed in 
the Appendix) and for three two-species fermions in 3D~\cite{Kartavtsev06}. 

\subsection{Asymptotic expansions and boundary conditions for HREs}
\label{Asymptotic}

Asymptotic expansions for all the terms of HREs are of interest for 
qualitative study of the described three-body system. 
In addition, the explicit asymptotic form allows one to formulate the boundary 
conditions and to improve the accuracy of the numerical calculations. 
The analytical expressions derived in the preceding sections provide
a straightforward determination of the eigenpotentials and the coupling terms
in the asymptotic region $|R| \to \infty$.

\begin{figure}[hbt]
\includegraphics[width=0.48\textwidth]{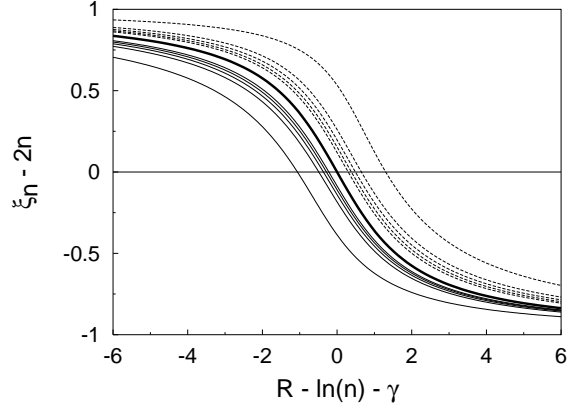}
\caption{Convergence of eigenvalues $\xi_n(R) $ to the limiting function 
(bold line). 
Two families of $\xi_n(R) $ are plotted by solid lines for $n = 3m$ and by 
dashed lines for $n = 3m + 1$, $m = 1, 5, 9, 15, 25$. 
\label{figxi1}}
\end{figure}

The expansion of eigenpotentials $\xi_n^2(R)$ at $|R| \to \infty$ follows from 
the expansion of the eigenvalue equation (\ref{transeq}) at the singular 
points, i.~e., near the odd integer $\xi$ (except $\xi = 3$) and at infinite 
$\xi$. 
In particular, the expansion at $\xi\to i\infty$ provides the lowest
eigenpotential at $R \to \infty$
\begin{eqnarray}
\label{as1+}
\xi^2_1(R) = -4e^{2(R - \gamma)} - \frac{1}{3} -
\frac{2}{45}e^{-2(R - \gamma)} + O(e^{-4R})\ .
\end{eqnarray}
The neighboring branches of the multivalued function $\xi(R)$ are
continuously connected at infinity so that $\xi_n(R)$ at $R \to \infty$ is
continuation of $\xi_{n-1}(R)$ at $R \to -\infty$.
Thus, the same asymptotic expansion at $R \to \infty$ for $\xi_n(R)$ and
$\xi_{n-1}(-R)$ is obtained by using the expansion of $R(\xi)$ near the odd
integer $\xi$,
\begin{equation}
\label{asn+}
\xi_n(R) = \xi_{n-1}(-R) = \left\{
\begin{array}{l}
1 + \displaystyle\frac{3}{R + \ln(4/3)} + O(|R|^{-3}) \ , \ n = 2 \\
2 n - 1 + \displaystyle\frac{1 - 2(-1)^n P_{n - 1}(1/2)}
{R - \gamma - \psi(n) + (-1)^n \frac{d P_\nu\left({1}/{2}\right)}{d\nu}
|_{\nu = n - 1} } + O(|R|^{-3}) \ , \ n > 2 \ .
\end{array}
\right.
\end{equation}
As $\xi_n(R)$ ($n \ge 2$) are of the smoothed-step form with the steepest 
descent at $R \approx \ln{n} $, the asymptotic expansion~(\ref{asn+}) is not 
uniform in $n$, viz., it is valid only if $R \gg \ln{n} $, which hinders any 
consideration of the infinite $n$ limit. 
Therefore, one needs the asymptotic expansion at $R \to \infty$ which 
reproduces the step-like dependence of $\xi_n(R)$ at least in the large-$n$ 
limit thus being applicable for both large $R$ and $n$. 
The expansion is constructed by using the requirement that both the $\xi_n(R)$ 
and their derivatives over $R$ for $n > 2$ coincide with the exact result at 
the point ${\bar R}_n = \gamma + \psi (n+1/2) - (-1)^{n}\pi P_{n - 1/2}(1/2)$, 
viz., one requires $\xi_n({\bar R}_n) = 2n $ and $\xi_n'({\bar R}_n) = 
-4 \left[ \pi^2 - 2\psi'(n + 1/2) + 
(-1)^n 2 \pi \frac{\partial P_{\nu}(1/2)}{\partial\nu}|_{\nu = n - 1/2} 
\right]^{-1}$, which leads to the result, 
\begin{equation}
\label{asnew}
\xi_n(R) \approx 2n + \frac{2}{\pi}\left[ \arctan{x_n} + (-1)^n
\arcsin\frac{\pi \xi'_n({\bar R}_n) P_{n - 1/2}(1/2)}{\sqrt{x_n^2 + 1}}\right],
\end{equation}
where $x_n = \displaystyle\frac{\pi}{2}\xi'_n({\bar R}_n)
\left[ R - \gamma - \psi (n + 1/2)\right] $.
As follows from~(\ref{asnew}), $\xi_n(R) $ (properly shifted along both 
coordinate axes) at large $n$ converge to the function, 
$\xi_n(R) \approx 2n - \displaystyle\frac{2}{\pi} 
\arctan{\frac{2}{\pi}(R - \ln{n} - \gamma )}$. 
The quite slow (as $n^{-1/2}$) large-$n$ convergence is entirely determined 
by the asymptotic behaviour of the Legendre function as $\nu \to \infty $, 
$P_{\nu - 1/2}(1/2) \sim \nu^{-1/2}\cos{\pi(\nu/3 - 1/4)}$ \cite{Bateman53}. 
Actually, the terms of order $\sim n^{-1/2}$ contain the dependence on $n$ via 
the expressions $(-1)^n \cos{\pi(n/3 - 1/4)}$ and 
$(-1)^n \sin{\pi(n/3 - 1/4)}$, which are the periodic functions of $n$ 
with period 3. 
Thus, one concludes that $\xi_n(R) $ up to the leading order terms in $n$
belong to three families for different $n$ mod 3. 
Convergence to the unique function is illustrated in Fig.~\ref{figxi1} for 
two families of $\xi_n(R)$. 

\begin{figure}[htb]
\includegraphics[width=0.48\textwidth]{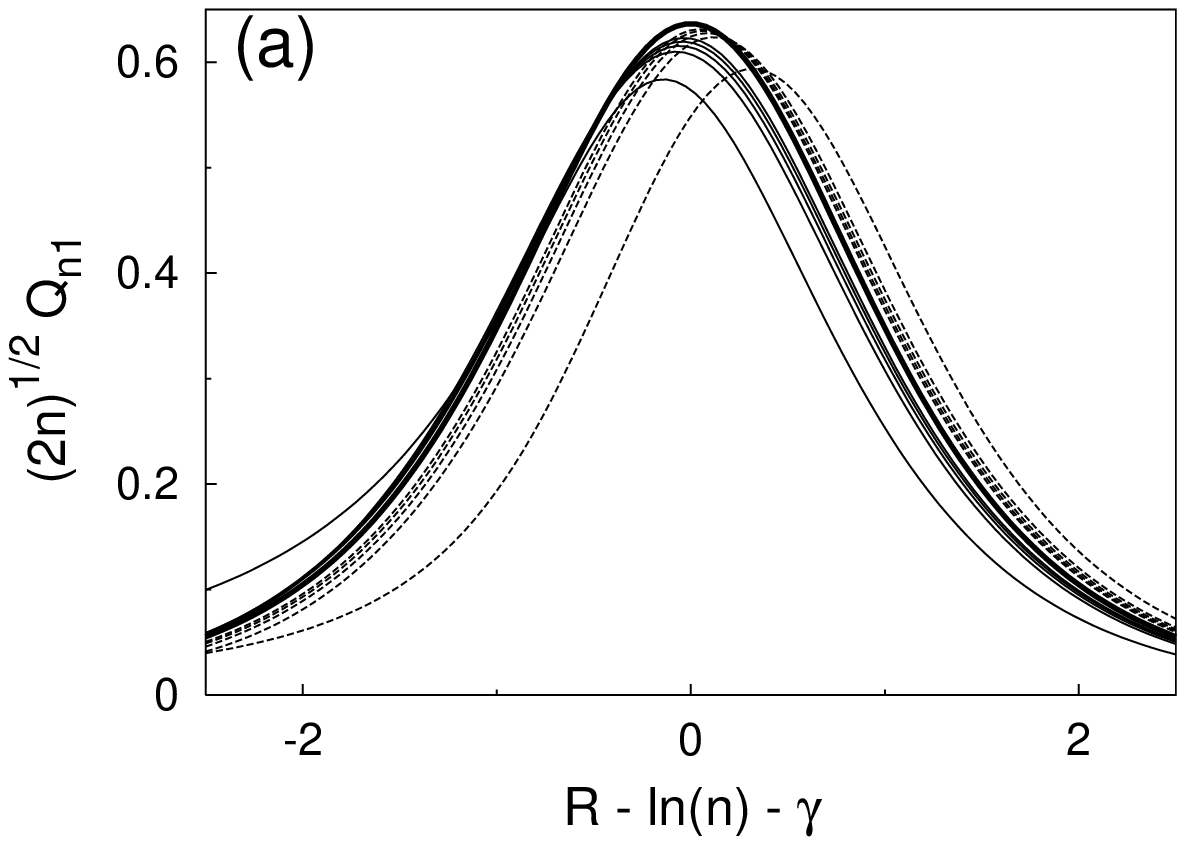}
\includegraphics[width=0.48\textwidth]{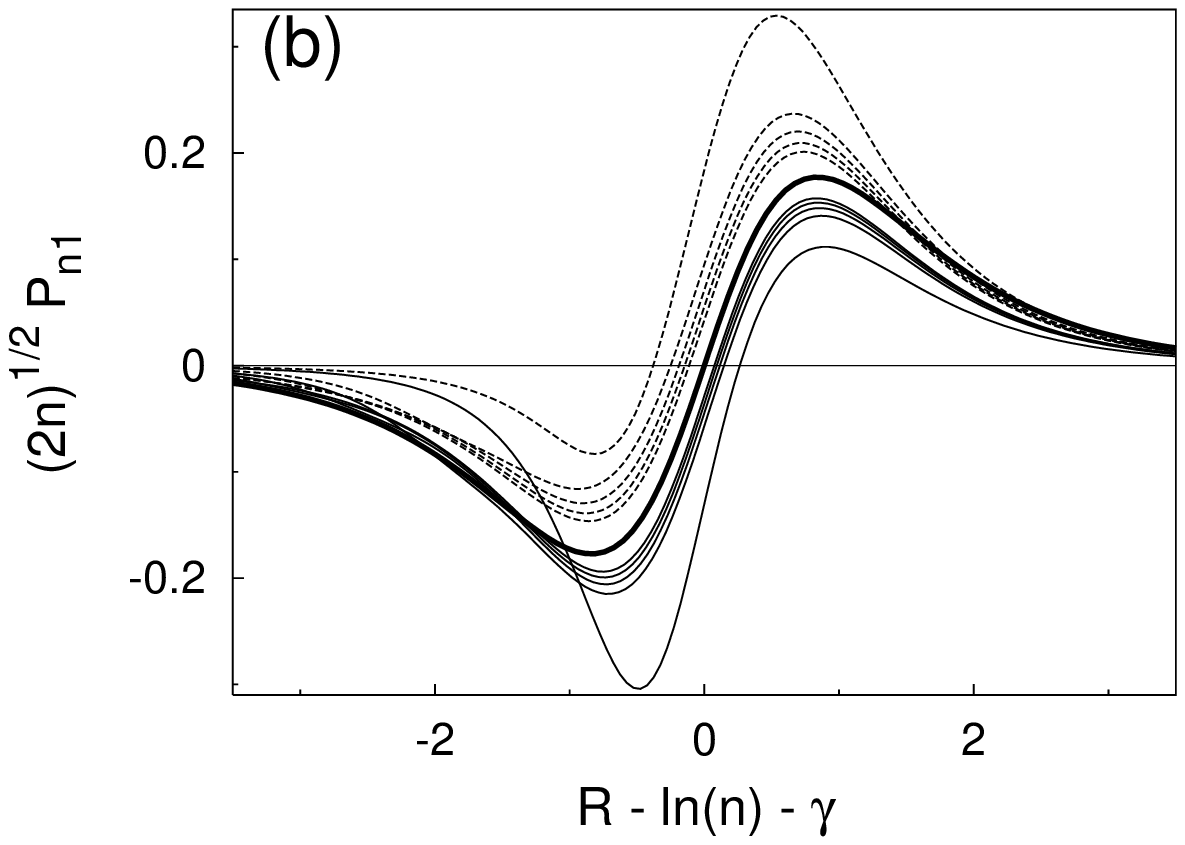}
\includegraphics[width=0.48\textwidth]{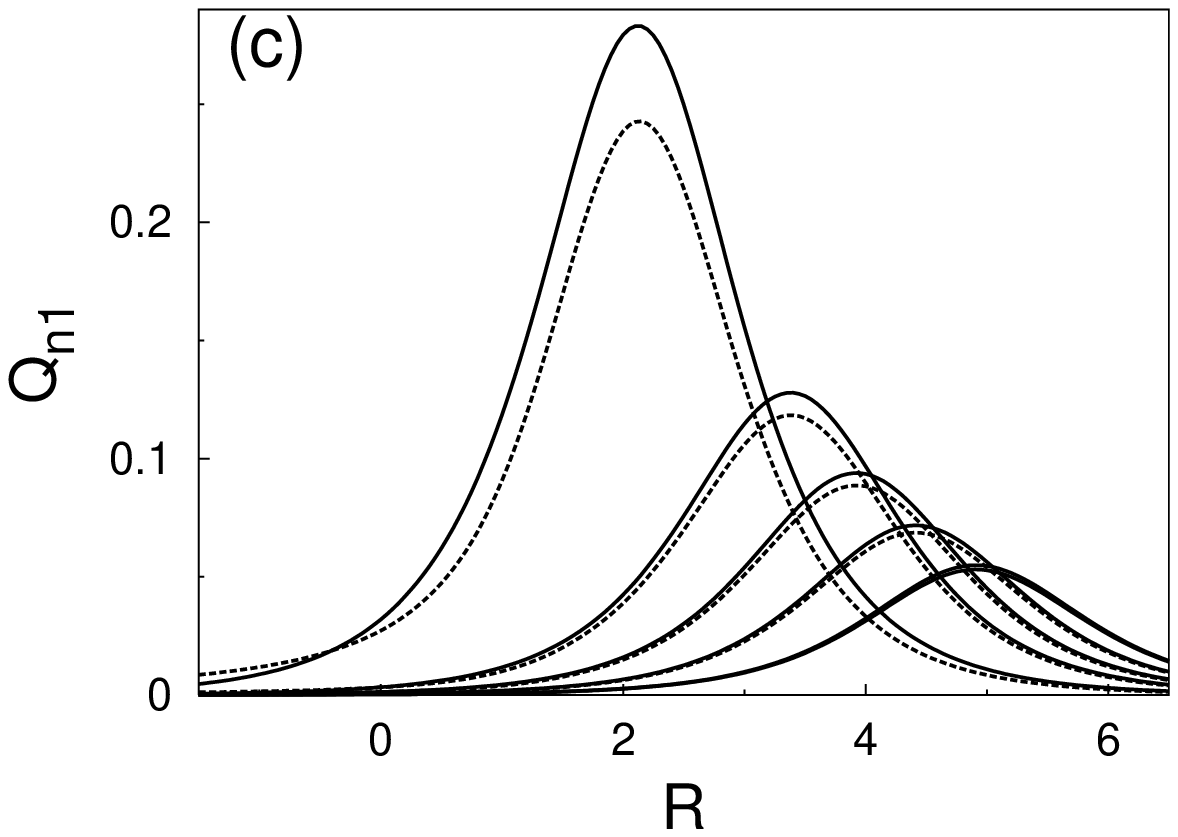}
\caption{Two families of $Q_{n1}(R)$ (a) and $P_{n1}(R)$ (b) are plotted by 
solid lines for $n = 3m $ and by dashed lines for $n = 3m + 1$, 
$m = 1, 5, 9, 15, 25$. 
The corresponding limiting functions ${\tilde Q}_1(R - \ln n - \gamma ) $ and
${\tilde P}_1(R - \ln n - \gamma ) $ are plotted by a bold line.
In panel (c) the exact result (solid lines) and the asymptotic expression 
(\protect\ref{asqq}) (dashed lines) are compared for the third family of 
$Q_{n1}(R)$ ($n = 3m + 2$, $m = 1, 5, 9, 15, 25$). 
\label{figq}}
\end{figure}

Substituting the above expansions for $\xi_n(R)$ in the analytical 
expressions derived in Section~\ref{Derivation}, one obtains asymptotic 
expansions of the coupling terms. 
A separate expression for the first-channel diagonal coupling term 
at $R \to \infty$ follows from~(\ref{as1+}), 
\begin{equation}
\label{p11as}
P_{11}(R) = {1}/{3} + 2/45 e^{-4(R - \gamma)} + O(e^{-6R}) \ . 
\end{equation}
Furthermore, using the expansions~(\ref{asn+}) one finds that $Q_{nm}(R)$ 
decrease as $|R|^{-2}$ and $P_{nm}(R)$ decrease as $|R|^{-4}$ except the terms 
$Q_{n1}(R)$ and $P_{n1}(R)$ at positive $R$, which decrease as 
$Q_{n1}(R) \sim P_{n1}(R) \sim e^{-R}R^{-1}$ at $R \to \infty$ 
provided $R \gg \ln{n} $. 
As discussed above, this asymptotic dependence is not uniform in $n$ and one 
would use the expression~(\ref{asnew})to obtain the uniform expansion which 
is valid for large $n$. 
For example, the desired expansion for $Q_{nm}(R)$ at $R \to \infty$ takes 
the form 
\begin{equation}
\label{asqq}
Q_{n1}(R) \approx \frac{\pi \xi'_n({\bar R}_n)}{4n}
\left[ \frac{\xi_n (R)}{x_n^2 + 1}
\left(1 - \frac{(-1)^n 2 x_n P_{n - 1/2}(1/2)}{\sqrt{x_n^2 + 1}}\right)
\right]^{1/2} 
\left[\cosh\left(\frac{2x_n}{\pi\xi'_n({\bar R}_n)}\right)\right]^{-1}  \ .
\end{equation}
Similar to $\xi_n(R)$, both $Q_{nm}(R)$ and $P_{nm}(R)$ (properly scaled
and shifted along the coordinate axis) converge at large $n$ to the universal
limiting functions so that $Q_{n1}(R) \to (2 n)^{-1/2}
{\tilde Q}_1(R - \ln n - \gamma ) $, $P_{n1}(R) \to (2 n)^{-1/2}
{\tilde P}_1(R - \ln n - \gamma ) $, $P_{nn}(R) \to
{\tilde P}(R - \ln n - \gamma ) $, and $Q_{nm}(R) \to
{\tilde Q}(n/m, R - \ln n - \gamma ) $ for $ n > m$, where
\begin{eqnarray}
\label{q1univ}
& & {\tilde Q}_1(y) = \left( y^2 + \pi^2/4 \right)^{-1/2} 
\left(\cosh{y}\right)^{-1} \ , \\
\label{p1univ}
& & {\tilde P}_1(y) = {\tilde Q}_1(y)\left(\tanh{y} - \frac{y}{y^2 +
\pi^2/4 } \right) \ , \\
\label{pduniv}
& & {\tilde P}(y) = \frac{\pi^2}{12 \left( y^2 + \pi^2/4 \right)^2}\ , \\
\label{quniv}
& & {\tilde Q}(z, y) = z^{1/2} \left\{ (z^2 - 1)\left( y^2 + \pi^2/4 \right)
\left[ (y + \ln{z})^2 + \pi^2/4 \right] \right\}^{-1/2} \ .
\end{eqnarray}
Splitting of eigenpotentials into three families depending on $n$ mod 3 
entails corresponding splitting of the coupling terms. 
The splitting and convergence to the universal limiting functions for 
$Q_{n1}(R)$ and $P_{n1}(R)$ are illustrated in Fig.~\ref{figq}. 

At infinite separation of particles, i.~e., when $R \to \infty$, the leading 
term of the asymptotic expansion for the first eigenpotential is related to 
the dimensionless energy of the two-body bound state so that 
$\xi^2_1 e^{-2R} = \xi^2_1/\rho^2 \to -4e^{-2\gamma} \approx -1.261$, while 
for other eigenpotentials the leading terms in the upper channels are related 
to the kinematic barriers, $\xi^2_2 e^{-2R} \to 1/\rho^2$ and 
$\xi^2_n e^{-2R} \to (2n - 1)^2/\rho^2$ for $n > 2$. 
Thus, in the asymptotic region, the first-channel component of the total wave 
function describes the two-cluster $2+1$ configuration, whereas 
the upper-channel components describe the three-cluster configuration. 

Using the expansion of $\xi_1(R)$ at $R \to -\infty$~(\ref{asn+}), one obtains 
the asymptotic form of the first-channel radial function at a small 
hyper-radius 
\begin{equation}
\label{f1as}
f_1(R) \sim e^R \left( R + \ln\frac{4}{3}\right)^2
\left( R + \ln\frac{4}{3} - \frac{3}{2} \right) \ .
\end{equation}
Given the expansion of $\xi_1(R)$, Eqs.~(\ref{faddeev}),~(\ref{chi}), 
(\ref{normconst}), and the expansion of the Legendre function at $\nu \to 0$, 
$P_{\nu}(-\cos 2\alpha) \approx 1 + 2 \nu \ln\sin\alpha $~\cite{Bateman53}, 
the asymptotic form of the first eigenfunction on the hypersphere 
at $R \to -\infty$ is 
\begin{equation}
\label{Phias}
\Phi_1(\alpha , R) \sim \left(R + \ln\frac{4}{3} \right)^{-2} 
\left( R + \ln\frac{4}{3} + \frac{3}{2} \right) 
\left(R + \ln\frac{4}{3} + \sum_i\ln\sin\alpha_i \right) \ . 
\end{equation}
As the first-channel contribution dominates in the series~(\ref{Psi}), 
the expressions~(\ref{f1as}),~(\ref{Phias}) entail the asymptotic form 
of the total wave function at $R \to -\infty$, i.~e., near the 
triple-collision point, 
\begin{equation}
\label{psiaszero}
\Psi \sim \left( R + \ln\frac{4}{3} \right)^2 
\left( \sum_i\ln\sin\alpha_i + R + \ln\frac{4}{3} \right) = 
\ln^2\frac{4}{3}\rho \ln\frac{4 x_1 x_2 x_3}{3 \rho^2} \ . 
\end{equation}
In addition, the non-singularity of the lowest eigenpotential in the limit of 
a small hyper-radius $R \to -\infty$ leads to the well-known conclusion that 
neither Efimov nor Thomas effects exist in 2D~\cite{Bruch79,Lim80,Nielsen99}. 

For the eigenvalue problem, i.~e., for calculation of the bound-state 
energies, the solutions satisfy the requirement of the square integrability of 
the total wave function, $\displaystyle\sum_n\int_{-\infty}^{\infty} 
f_n^2(R)e^{2R}dR = 1$, and in practice one can use the boundary conditions for 
the channel functions of the form $f_n(R) \to 0 $ at $R \to \pm\infty$. 
The asymptotic boundary conditions for the low-energy scattering of the third 
particle off the bound pair are similar to those for the two-body scattering. 
Below the three-body threshold, the wave function in the asymptotic region 
tends to a product of the two-body bound-state wave function $\varphi (x)$ 
and the function $F(r)$, which depends on the inter-cluster distance 
$r = \sqrt{3}y/2 $ and describes relative motion of the third particle and 
a bound pair. 
At the threshold, i.~e., at the zero kinetic energy of colliding particles, 
the 2 + 1 scattering length $A$ is defined by the two-cluster asymptotic form 
$F(r) \sim \ln(r/A)$, which leads to the expression $\Psi ({\mathbf x}, 
{\mathbf y}) \sim \varphi(x) \ln\displaystyle\frac{\sqrt{3}y}{2A}$ at 
$y \to \infty $. 
Taking into account that the first-channel eigenfunction 
$\Phi_1(\alpha,\theta, R)$ at a large hyper-radius reduces to $\varphi(x)e^R$ 
and $y\approx\rho = e^{R}$ for $x\ll y$, one finds the asymptotic form 
of the channel function $f_1(R)$,
\begin{eqnarray}
\label{A3bound1}
f_1(R) \sim \ln\frac{2A}{\sqrt{3}} - R\ ,\qquad R \to \infty\ .
\end{eqnarray}
In addition to the asymptotic expression~(\ref{A3bound1}) at $R \to \infty$, 
the first channel function $f_1(R) \to 0$ at $R \to -\infty$, while all other 
channel functions $f_n(R) \to 0$, $n \geq 2$ at both limits 
$R \to \pm \infty$. 

The asymptotic form of $\xi_1^2(R)$ and $P_{11}(R)$ at $R \to \infty$ are of 
fundamental importance in the analysis of the low-energy $ 2 + 1 $ scattering. 
In the lowest channel of HREs, the leading term of $\xi_1^2(R)$~(\ref{as1+}) 
cancels the term $E e^{2R}$ for the threshold energy $E = -4e^{-2\gamma}$ and 
the next-order constant $1/3$ terms of $\xi_1^2(R)$~(\ref{as1+}) and 
$P_{11}(R)$~(\ref{p11as}) cancel each other, therefore, the effective 
interaction takes the form $\displaystyle\frac{2}{45}e^{-2(R - \gamma)}$, 
which corresponds to the polarization interaction $V_p = -\alpha / (2 r^4) $, 
where $r$ is the distance between a dimer and the third particle and 
$\alpha = e^{2\gamma }/ 20 \approx 0.1586$. 
This long-range polarization tail of the effective interaction is a specific 
2D feature (compare, e.~g., the exponential fall-off of the lowest effective 
interaction at large distances for three bosons in 3D~\cite{Kartavtsev99}). 
The $2 + 1$ scattering length in 2D exists even if the effective interaction 
contains the polarization tail~\cite{Bolle84,Verhaar85} (in fact, for the 
potentials decreasing faster than $1/\rho^{2 + \delta}$). 
This can be seen from the asymptotic solution of the first-channel HRE 
of the system~(\ref{system1}) at the threshold energy, $E = -4e^{-2\gamma}$. 
Up to terms of order $O(e^{-2R})$, the first-channel HRE takes the form, 
\begin{equation}
\label{hre1as}
\left[\frac{d^2}{d R^2} + {2 \over 45}e^{-2(R - \gamma )}\right]f_1(R) = 0 \ , 
\end{equation}
the general solution of~(\ref{hre1as}) is the linear combination of 
the Bessel functions 
\begin{eqnarray}
\label{A3bound2}
f_1(R) \sim C_1 J_0\left(\sqrt{2/45}\ e^{-R + \gamma}\right) + 
C_2 Y_0 \left(\sqrt{2/45}\ e^{-R + \gamma}\right) \ . 
\end{eqnarray}
The asymptotic expansion of the solution~(\ref{A3bound2}) at $R \to \infty$, 
$f_1(R) \sim \displaystyle\frac{\pi}{2} \frac{C_1}{C_2} + 2\gamma - 
\frac{1}{2}\ln90 - R $, is of the form~(\ref{A3bound1}), which proves 
the existence of the scattering length $A$. 
As a consequence, the leading-order terms of the effective-range expansion for 
$2 + 1$ scattering are of the usual form~(\ref{f0}) for two-body scattering in 
2D, viz., $\displaystyle\frac{\pi}{2}\cot\delta (k) \approx \ln (k A/2) + 
\gamma $, where $k$ is the wave number for the relative motion of a dimer and 
the third particle.
Nevertheless, the higher terms of the effective-range expansion 
are modified by the polarization tail of the effective interaction 
as is known to be the case in 3D scattering~\cite{OMalley61}. 

The role of the long-range term $\sim e^{-2R}$ or $\sim \rho^{-4}$ in 
the first-channel HRE requires a careful treatment because there is no clear 
reason for appearance of the polarization potential between a particle and 
a bound pair. 
In this respect, it is necessary to study a contribution of the upper channels 
to the effective dimer-particle interactions at long distances. 
Coupling with the upper channels produces in the first channel the nonlocal 
effective potential $U_c(R, R')$ which can be estimated in the lowest order 
of perturbation theory as $U_c(R, R') = \displaystyle\sum_n^{\infty} F_n(R) 
g_n(R - R') F_n(R')$, where $F_n(R) = Q_{n1}(R) \displaystyle\frac{d}{dR} + 
\frac{d}{dR} Q_{n1}(R) + P_{n1}(R) $ and $g_n(R - R')$ is Green's function 
in the $n$th channel. 
Taking into account that $\xi_n^2(R) \sim 4 n^2 $, $Q_{n1}(R) = 
(2 n)^{-1/2} {\tilde Q}_1(R - \ln n - \gamma ) $, and $P_{n1}(R) = 
(2 n)^{-1/2} {\tilde P}_1(R - \ln n - \gamma ) $ for large $n$, one can 
estimate $g_n(x) \sim (4 n)^{-1} e^{-2 n |x|}$ and $F_n(R) \sim n^{-1/2} 
{\tilde F}(R - \ln{n}) $, where ${\tilde F}(x)$ is expressed via 
${\tilde Q}_1(x) $ and ${\tilde P}_1(x)$. 
As $g_n(x) \to (2 n)^{-2} \delta (x)$ for $n \to \infty$, these estimates 
entail the following local limit of $U_c(R, R')$, viz., 
$U_c(R) \sim \displaystyle\sum_n^{\infty} n^{-3} {\tilde F}^2(R - \ln{n}) $. 
Summing over $n$, one finds that the leading term of the effective 
potential is $U_{c}(R) \sim e^{-2R}$, in other words, coupling with the upper 
channels produces in the first channel the long-range term of the same order 
$\sim e^{-2R}$ or $ \sim \rho^{-4}$ as the above-discussed polarization tail. 
Thus, any conclusion on the long-range behaviour of the wave function or, 
equivalently, on the next-to-leading terms of the effective-range expansion 
for $2 + 1$ scattering must be based on the study of a large number of HREs. 

\section{Numerical calculations}
\label{Numerical}

The eigenpotentials $\xi_n^2(R)$ and the coupling terms $P_{nm}(R)$ and 
$Q_{nm}(R)$ in HREs were calculated by solving the transcendental eigenvalue 
equation~(\ref{transeq}) and by using Eqs.~(\ref{Qn3}),~(\ref{Pn2}) 
and (\ref{Pn4}). 
The derivatives with respect to $R$ ($\xi'_n$, $\xi''_n$ and $\xi'''_n$) were 
replaced by the derivatives of the inverse function ($dR/d\xi$, $d^2R/d\xi^2$, 
and $d^3R/d\xi^3$) which are easily calculable from the eigenvalue 
equation~(\ref{transeq}). 
The most involved numerical problem is to calculate the Legendre function 
and its derivatives with respect to the index entering into 
Eqs.~(\ref{transeq}),~(\ref{Qn3}), (\ref{Pn2}) and (\ref{Pn4}).
This is done for both real and imaginary $\xi$ by using the Mehler-Dirichlet 
integral representation~\cite{Bateman53}, 
\begin{equation}
\label{Pmd}
P_{\frac{\xi-1}{2}} \left(\frac{1}{2}\right) = \frac{\sqrt{2}}{\pi} 
\int\limits_{0}^{\pi /3} \frac{d t\cos{\frac{\xi}{2}t }} 
{\sqrt{\cos{t}-1/2}} \ , 
\end{equation}
for the Legendre function and using for its derivatives the corresponding 
integral representations obtained by differentiating Eq.~(\ref{Pmd}) with 
respect to $\xi$. 
The terms containing an integrable square-root singularity are subtracted from 
the integrand and calculated exactly to improve the accuracy. 
As a result, the Legendre function was calculated with a relative accuracy 
about $10^{-11}$ whereas the accuracy degraded about one order for each of 
the subsequent derivatives. 
\begin{figure}[thb]
\includegraphics[height=.25\textheight,width=0.4\textwidth]{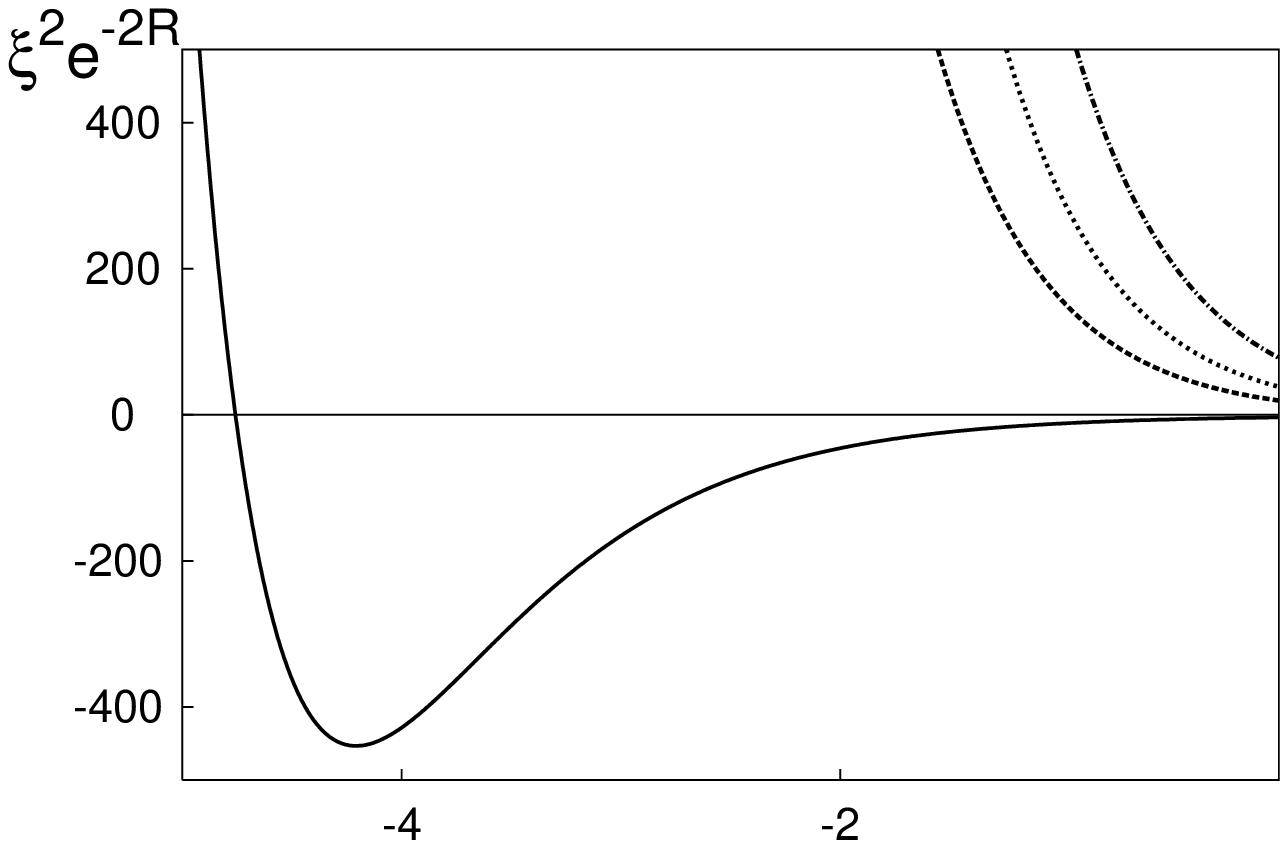}
\hskip-.95cm
\includegraphics[height=.25\textheight,width=0.4\textwidth]{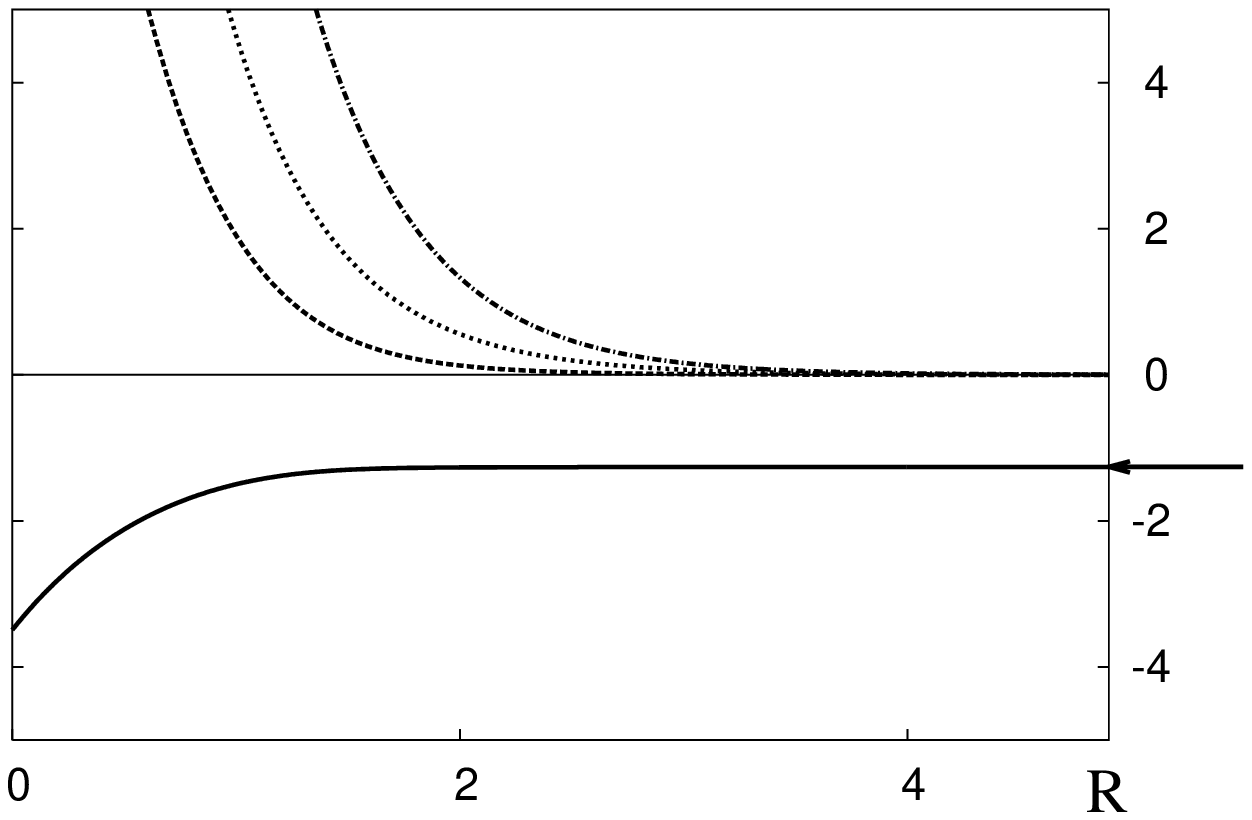}
\\
\caption{ The lowest scaled eigenpotentials $\xi^2_n(R)e^{-2R}$.
Notice different scales for the positive and negative $R$.
The arrow marks the two-body bound-state energy $-4e^{-2\gamma}$.
\label{fig1}}
\end{figure}
As mentioned in Section~\ref{Asymptotic}, accuracy of the numerical 
calculation suffers from the subtraction of divergent terms in the vicinity of 
the exceptional point $\xi = 3$. 
For this reason, $\xi_2(R)$, $Q_{2n}(R)$, and $P_{2n}(R)$ in a narrow region  
around the point $R_0 \approx 1.64$ (which corresponds to $\xi_2(R_0) = 3$) 
were obtained by the interpolation procedure. 
Under the described approximations, the overall relative accuracy was not 
worse than $10^{-11}$ for the eigenpotentials and $10^{-8}$ for the coupling 
terms. 
It is worthwhile to mention that less accurate calculation of the coupling 
terms is in accordance with a smaller contribution of these terms to the final 
values. 
The sum rule~(\ref{Pprop1}) for the coupling terms was numerically checked 
and it was found that the difference $\sum_{k=1}^{N}Q_{nk}Q_{mk} - P_{nm}$ 
decreases as $N^{-2}$ with increasing $N$. 
The eigenpotentials and all the coupling terms for the four lowest channels of 
HREs are shown in Figs.~\ref{fig1},~\ref{fig2}. 
\begin{figure}[thb]
\includegraphics[width=0.48\textwidth]{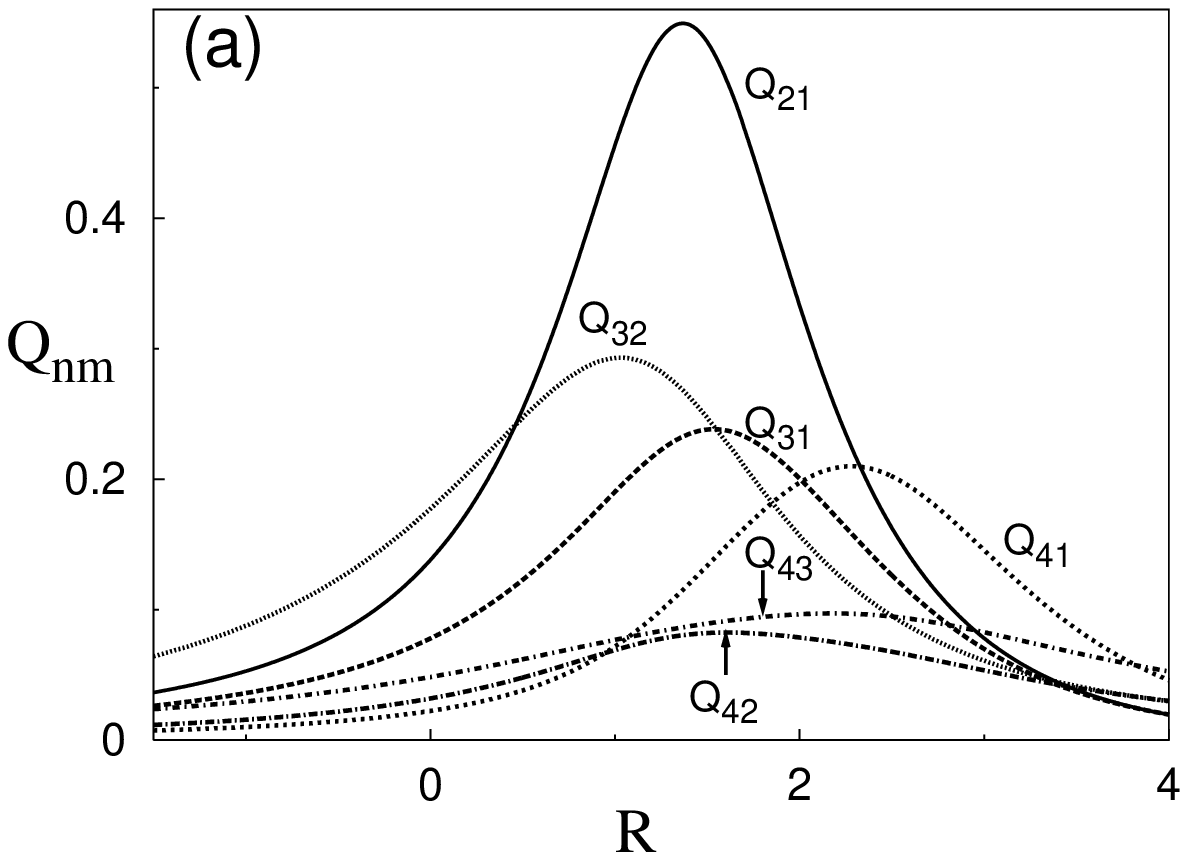}
\hskip-0.5cm
\includegraphics[width=0.48\textwidth]{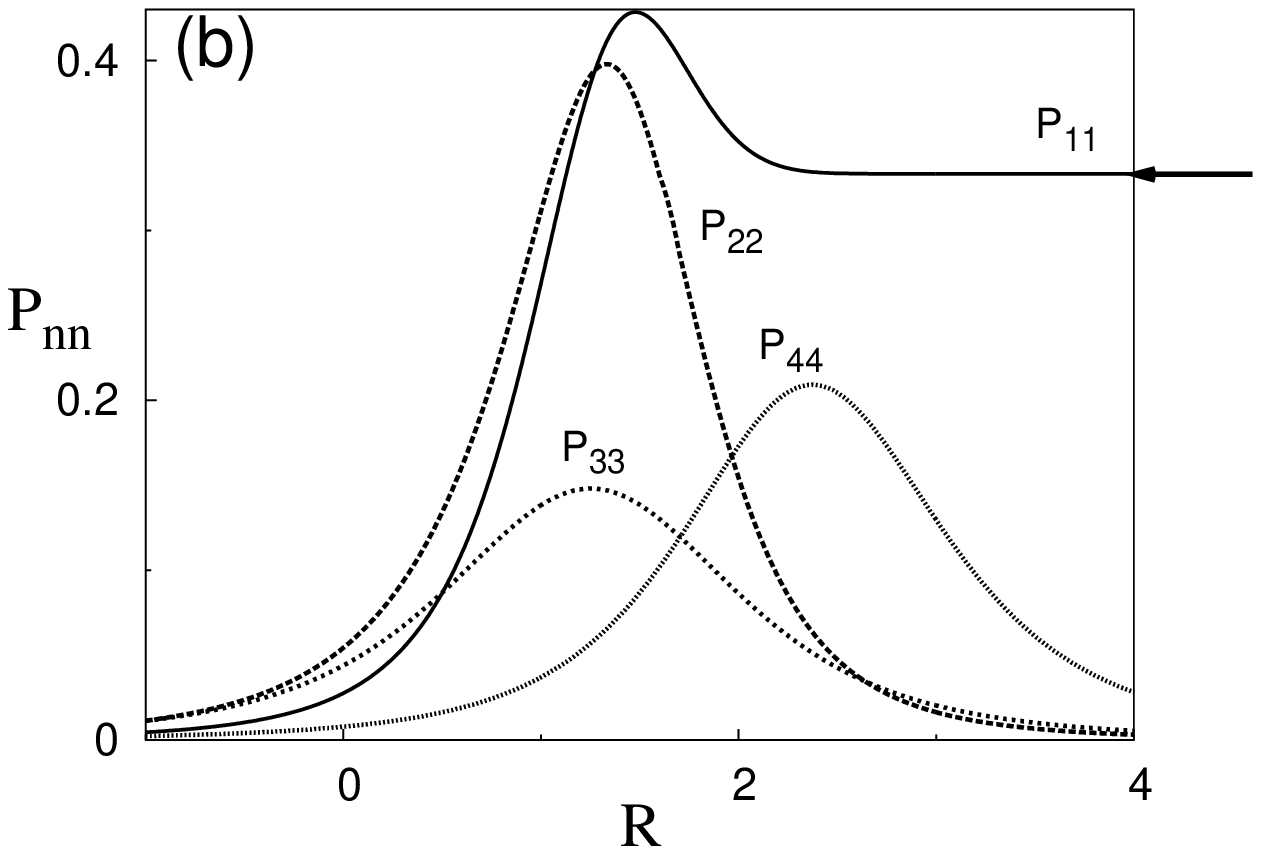}
\hskip-0.5cm
\includegraphics[width=0.48\textwidth]{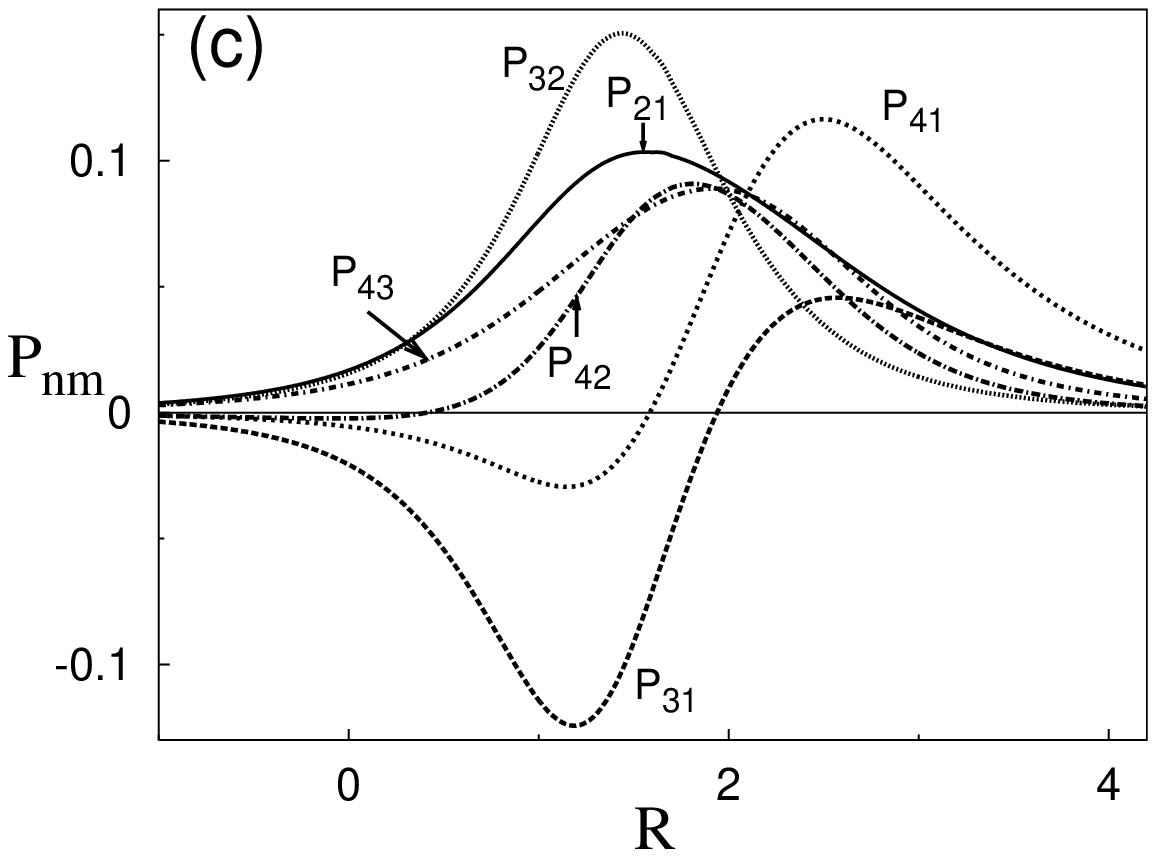}\\
\caption{Coupling terms $Q_{mn}(R)$ (a), $P_{nn}(R)$ (b), and $P_{nm}(R) $
for $n \ne m$ (c). The arrow marks the large hyper-radius limit $1/3$ of
$P_{11}(R)$.
\label{fig2}}
\end{figure}

For numerical solution, the truncated system of $N$ HREs is reduced to 
the form without the first derivatives by the transformation 
${\mathrm f}(R) = {\mathrm T}(R)\tilde{{\mathrm f}}(R)$, where the orthogonal 
matrix ${\mathrm T}(R)$ satisfies the equation 
\begin{eqnarray}
\label{T}
\frac{d {\mathrm T}}{d R} + {\mathrm Q}{\mathrm T} = 0\ . 
\end{eqnarray}
Furthermore, one introduces the antisymmetric matrix ${\mathrm B}$ by 
the Cayley transform, 
${\mathrm B} = ({\mathrm T} - 1)({\mathrm T} + 1)^{-1}$, and solves 
the equation 
\begin{eqnarray}
\label{eqB}
2\frac{d {\mathrm B}}{d R} = \left( {\mathrm B} - 1 \right) {\mathrm Q} 
\left( {\mathrm B} + 1 \right) \ . 
\end{eqnarray}
This form is preferable because one can use only the upper triangle of 
the matrices ${\mathrm B}$ and ${\mathrm Q}$ in the numerical calculations, 
which gives the antisymmetric matrix ${\mathrm B}$ and the orthogonal matrix 
${\mathrm T} = (1 - {\mathrm B})(1 + {\mathrm B})^{-1}$ independently of 
the round-off error. 
Note that in the two-channel approximation the non-zero matrix elements of 
${\mathrm B}$ are explicitly expressed via the quadrature, 
$B_{21} = -B_{12} = \tan\frac{1}{2}\int\limits Q_{12}(R) d R $. 

Following the described procedure, the truncated system of $N$ HREs in two 
forms~(\ref{system1}) and~(\ref{system2}) was numerically solved on 
the finite interval [$R_{min}$, $R_{max}$]. 
At the first step, Eq.~(\ref{eqB}) was integrated and the matrix 
${\mathrm T} = (1 - {\mathrm B})^{-1}(1 + {\mathrm B})$ was determined at 
the mesh points on [$R_{min}$, $R_{max}$]. 
An arbitrary antisymmetric matrix ${\mathrm B}_0$ serves as the initial 
condition for the matrix equation~(\ref{eqB}) imposed at $R_{min}$. 
The consistency of the numerical procedure was additionally shown by checking 
the stability of the calculated values for different choices of the initial 
matrix ${\mathrm T}(R_{min})$. 
Given the calculated transformation matrix ${\mathrm T}(R)$, two eigenenergies 
and the scattering length were calculated by solving the eigenvalue problem 
and the scattering problem at the threshold energy $E = -4e^{-2\gamma}$ for 
the transformed HREs. 
The zero boundary conditions are imposed in the upper channels, i.~e., 
$f_n(R_{min}) = f_n(R_{max}) = 0$ for $n \ge 2$, whereas the left-end 
boundary condition in the first channel was determined from the asymptotic 
form of the $f_1(R)$ at $R \to -\infty$~(\ref{f1as}). 
At the right boundary, one uses $f_1(R_{max}) = 0$ for the eigenvalue problem 
and the asymptotic form~(\ref{A3bound2}) for the scattering problem. 
In the latter case, the scattering length is determined via the coefficients 
$C_{1, 2}$ calculated at $R_{max}$ , viz., $\ln{A} = 
\displaystyle\frac{\pi}{2}\frac{C_1}{C_2} + 2\gamma - \frac{1}{2}\ln{120} $, 
thus taking into account the polarization tail of the effective interaction 
beyond the integration region. 
The boundary conditions for the vector-function $\tilde{f}(R)$ were obtained 
by applying the transformation $T(R)$ at the points $R_{min}$ and $R_{max}$. 

\begin{figure}[htb]
\includegraphics[width = 0.48\textwidth]{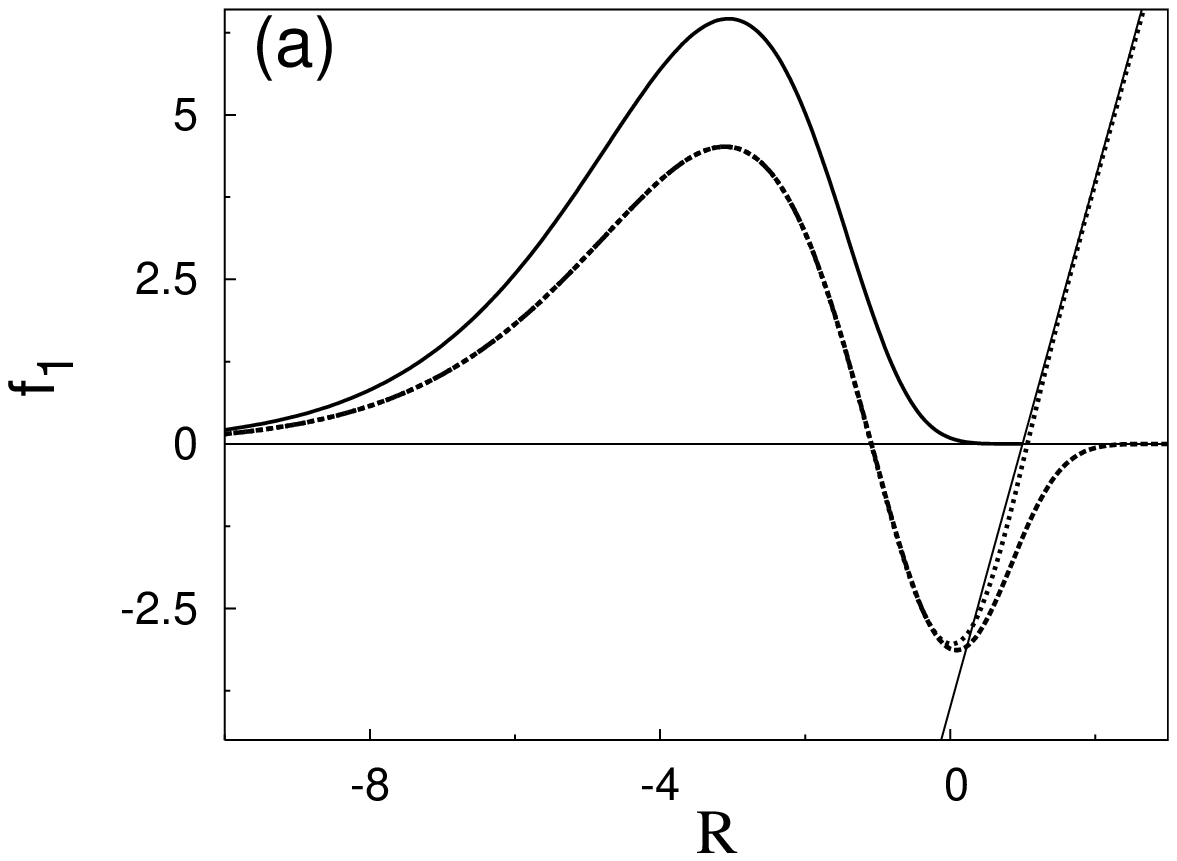}
\includegraphics[width = 0.48\textwidth]{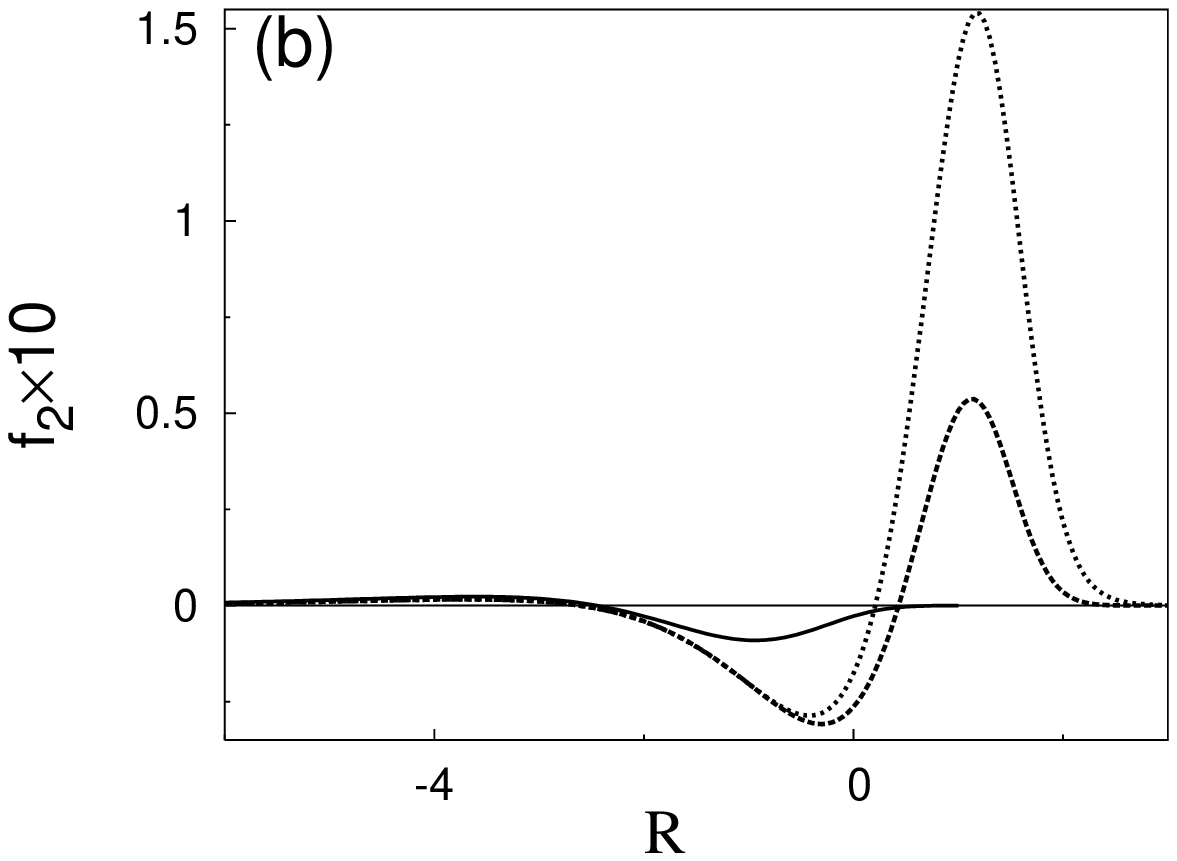}\\
\includegraphics[width = 0.48\textwidth]{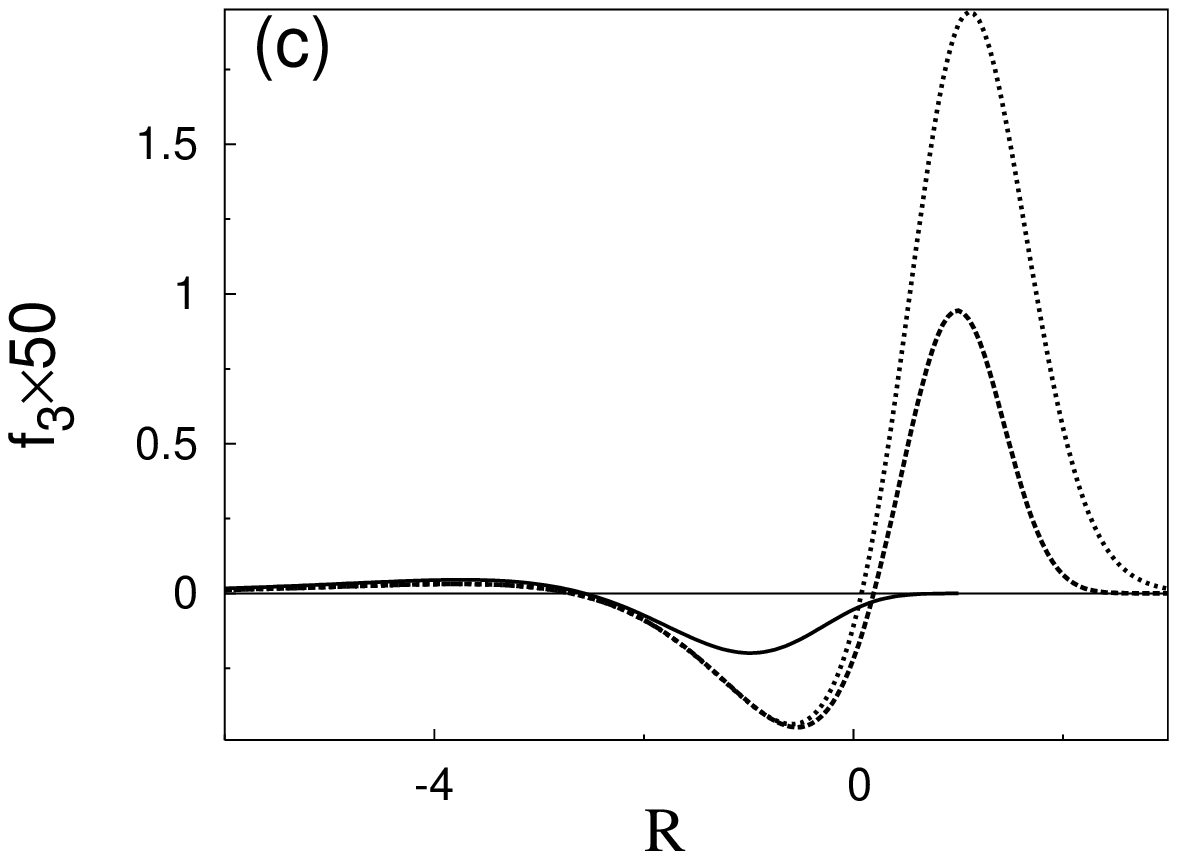}
\includegraphics[width = 0.48\textwidth]{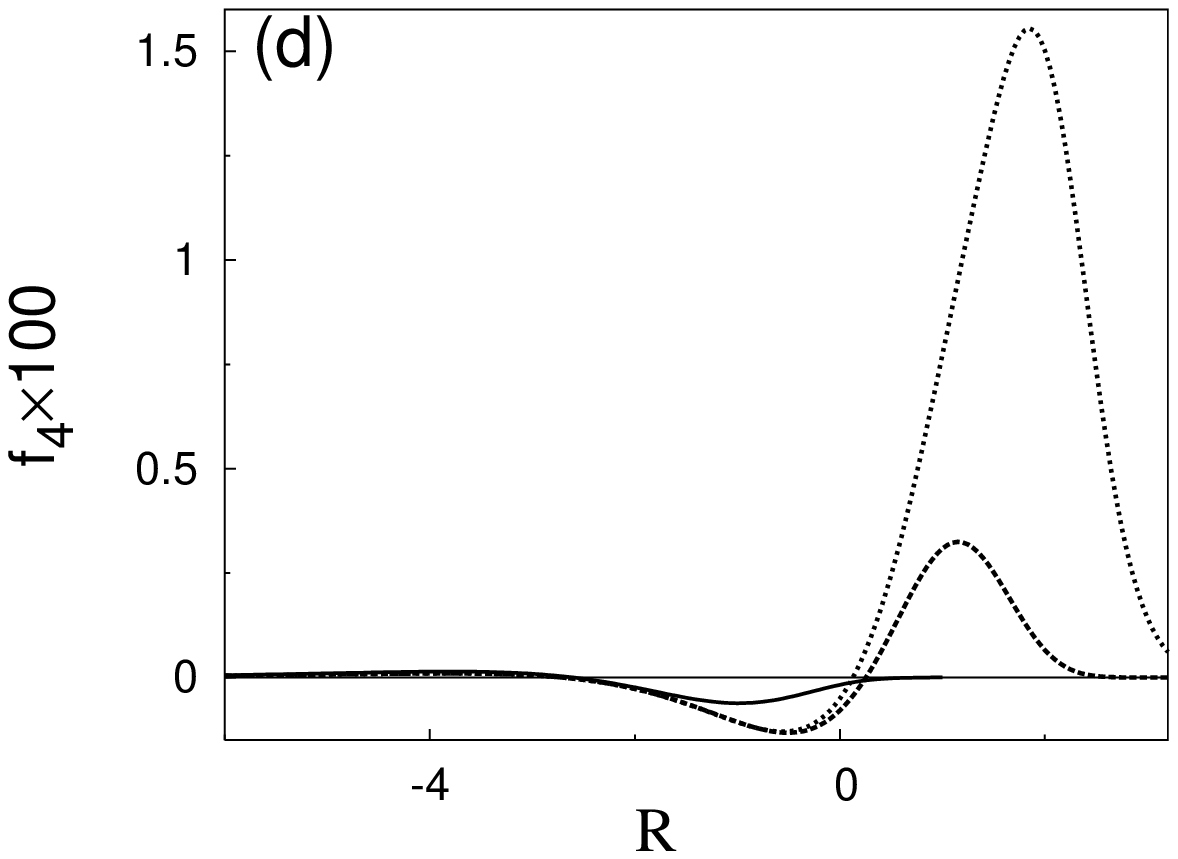}\\
{\caption{Radial functions of the four lowest channels $f_1(R)$ (a), $f_2(R)$  
(b), $f_3(R)$ (c), and $f_4(R)$ (d) for the ground state (solid lines), 
the excited state (dashed lines), and the scattering state calculated at 
the two-body threshold energy (dotted lines). 
For convenience, the radial functions for the excited state are multiplied by 
a factor 5, while those for the scattering state are scaled to match at 
the first maximum the first-channel functions of the excited state and 
scattering state. 
Linear asymptotic dependence of the first-channel scattering solution is 
shown by a thin straight line in panel (a).} 
\label{fig3}}
\end{figure}

The overall accuracy of the numerical procedure is estimated to provide 
the calculation of the binding energies and the scattering length with 
the relative error about $3 \cdot 10^{-8}$ and $1 \cdot 10^{-6}$, 
respectively. 
In particular, a sufficient accuracy of numerical integration of HREs was 
obtained by taking $R_{min} = -14$ and $R_{max} = 1.5$, $3.5$, and $6.0$ 
for the ground-state, the excited-state, and the scattering-length 
calculations, respectively. 
The structure of the calculated wave function is illustrated in 
Fig.~\ref{fig3}, where the four lowest channel functions $f_n(R)$ for  
the ground state, excited state, and the scattering state are shown. 
For convenience, the solution of the scattering problem is normalized 
to match the first-channel functions of the excited and scattering states at 
the point $R\approx -3.1$ corresponding to the first maximum. 
The numerical solution of the truncated system of $N$ HREs provides a set of 
binding energies and scattering lengths, which are presented in 
Table~\ref{tab1} in comparison with 
the calculations~\cite{Bruch79,Nielsen97,Hammer04}. 
\begin{table}[htb]
\caption{The three-body binding energies $\varepsilon_0$ and $\varepsilon_1$ 
(in units of the two-body binding energy) and the logarithm of the $2+1$ 
scattering length $A$ for identical bosons in 1D. 
The number of HREs is denoted by $N$ and the superscripts U and L mark 
the results obtained by solving HREs of the form~(\ref{system1}) 
and~(\ref{system2}), respectively. 
Shown are also the results of fitting the dependence on $N$ for the calculated 
binding energies and scattering length and those of other calculations.} 
\label{tab1}
\begin{tabular}{c@{\hspace{2mm}}cccccc@{\hspace{2mm}}c}
 $N$ & $\varepsilon_0^U$ & $\varepsilon_0^L$ & & $\varepsilon_1^U$ &
$\varepsilon_1^L$ & & $\ln{A} $ \\
$ 1 $ & $16.5194096$& $16.5788727$&& $1.26667318$& $1.29214773$&& $0.891305$\\
$ 2 $ & $16.5219444$& $16.5482471$&& $1.26998847$& $1.27658964$&& $0.858228$\\
$ 3 $ & $16.5226064$& $16.5302069$&& $1.27033831$& $1.27263368$&& $0.853238$\\
$ 4 $ & $16.5226348$& $16.5287316$&& $1.27036317$& $1.27217992$&& $0.851835$\\
$ 5 $ & $16.5226618$& $16.5267981$&& $1.27039042$& $1.27147416$&& $0.849801$\\
$ 6 $ & $16.5226787$& $16.5249848$&& $1.27040205$& $1.27101864$&& $0.848804$\\
$ 7 $ & $16.5226811$& $16.5246296$&& $1.27040405$& $1.27091797$&& $0.848343$\\
$ 8 $ & $16.5226835$& $16.5241930$&& $1.27040625$& $1.27077981$&& $0.847726$\\
$ 9 $ & $16.5226854$& $16.5237285$&& $1.27040762$& $1.27066543$&& $0.847341$\\
$10 $ & $16.5226859$& $16.5235979$&& $1.27040796$& $1.27063093$&& $0.847125$\\
$12 $ & $16.5226867$& $16.5232644$&& $1.27040864$& $1.27054373$&& $0.846651$\\
$14 $ & $16.5226870$& $16.5231314$&& $1.27040883$& $1.27050912$&& $0.846376$\\
$16 $ & $16.5226871$& $16.5230155$&& $1.27040895$& $1.27048209$&& $0.846186$\\
$\infty$ & $16.5226874$& - && $1.27040911$& - && $0.8451 $\\
Ref.~\cite{Bruch79} & \multicolumn{2}{c}{$16.1\pm 0.2$} & &
\multicolumn{2}{c}{$1.25\pm0.05$} & & - \\
Ref.~\cite{Nielsen97} & \multicolumn{2}{c}{16.52} & &
\multicolumn{2}{c}{1.267} & & - \\
Ref.~\cite{Hammer04} & \multicolumn{2}{c}{16.522688} & &
\multicolumn{2}{c}{1.2704091} & & - \\
Ref.~\cite{Adhikari93} & \multicolumn{2}{c}{-} & & \multicolumn{2}{c}{-} & &
$\approx 1.1 $
\end{tabular}
\end{table}
It is clearly seen that highly accurate results can be obtained by means of 
the few-channel calculation of the form~(\ref{system1}). 
The contribution to the binding energies from the upper channels 
(for $N \ge 16$) turns out to be comparable with the numerical accuracy. 
The role of the upper channels can be estimated by fitting to the simple power 
dependence on $N$, which is routinely used in the variational calculations. 
In the present calculations, it is reasonable to fit separately each of three 
families, i.~e., to take into account the periodic dependence on $N$ for 
$N = 3m, 3m + 1, 3m + 2$. 
The calculated binding energies are fairly well fitted to the $a + b/N^c$ 
dependence for each family with the fitted value of power $c \approx 4$. 
The logarithm of the scattering length $\ln A$ converges slower with 
increasing $N$ than the binding energies, which is manifested by the smaller 
fitted power $c \approx 1 - 1.3$. 
The fitted binding energies and the scattering length corresponding to 
$N = \infty $ are presented in Table~\ref{tab1} with the overall fitting error 
in the last digit. 
As expected, the  solution of the truncated system~(\ref{system2}) provides 
slower convergence with increasing number of channels $N$ than those 
of the form~(\ref{system1}). 
The calculation based on the solution of the truncated system~(\ref{system2}) 
gives a set of binding energies converging as $N^{-2}$, which is connected 
with the corresponding convergence rate of $\sum_{k=1}^{N}Q_{nk}Q_{mk}$ 
to $P_{nm}$. 

The calculated binding energies coincide within the declared accuracy with 
the solution of the momentum-space integral equations~\cite{Hammer04}, which 
underlines equivalence of quite different approaches. 
The binding energies of a limited accuracy obtained by solving the system 
of HREs Ref.~\cite{Nielsen97} are in agreement with the one-channel 
calculation of the present paper. 
The older results of~\cite{Bruch79} obtained by solving the integral equations 
are of low accuracy; in addition, the ground-state energy of~\cite{Bruch79} is 
above the upper bound found in the present paper. 
The calculations of the $2 + 1$ scattering length are rarely available in 
the literature. 
The present calculation of the $2 + 1$ scattering length in the universal 
limit could be compared with the results of Ref.~\cite{Adhikari93} by 
analyzing the dependence of $\bar{a}_3$ on $\bar{a}_2$ shown in Fig.~1 of that 
paper. 
The three-boson scattering length $\bar{a}_3$ is related to the scattering 
length $A$ defined in the present paper as 
$\bar{a}_3 = (2/\pi )\ln (2A/\sqrt{3})$, whereas the two-body scattering 
length $\bar{a}_ 2$ is defined in~\cite{Adhikari93} so that the universal 
limit corresponds to $\bar{a}_ 2 \to 0$. 
Considering the smallest $\bar{a}_2\approx 10^{-3}$ presented as the leftmost 
point in Fig. 1 of Ref.~\cite{Adhikari93}, one obtains $\bar{a}_3\approx 0.8$, 
i.e., $\ln A \approx 1.1$, which is well above the upper bound 
$\ln A \approx 0.8451$ calculated in the present paper. 
The discrepancy is presumably because the result of Ref.~\cite{Adhikari93} 
is not close enough to the universal limit and this points to the strong 
dependence $\bar{a}_3(\bar{a}_2)$ at $\bar{a}_ 2 \to 0$.

\section{Summary and discussion}

Universal description of three identical spinless bosons in 2D at low energy 
is expected by analogy with low-energy properties of two particles, which are 
universal (irrespective of a particular shape of the short-range potential) 
and parameterless if the only significant parameter, e.~g., the two-body 
scattering length $a$, is chosen as a scale. 
The two-body input completely determines the solution near 
the triple-collision point in the limit of the zero-range interactions, 
therefore, contrary to the corresponding problem in 3D, an additional 
regularization parameter is not necessary and there are neither Thomas nor 
Efimov effect in 2D.  
For this reason, a completely universal parameterless description exists in 
the low-energy limit and both three-body binding energies and the $2 + 1$ 
scattering length are the universal constants to be determined. 

The BCM is used to describe the pair-wise short-range interaction in 
the zero-range limit. 
The total wave function is expanded in a set of eigenfunctions on 
the hypersphere, which leads to a system of coupled HREs. 
The important point is that the analytical expressions are derived for all 
the terms of HREs, which allows one to study the asymptotic behaviour and 
to improve the accuracy of the numerical calculations. 
One should emphasize that the derivation is essentially based on application 
of the BCM and the Hellmann-Feynman-type relations, the latter are known to be 
useful in the calculation of the coupling terms~\cite{Stolyarov01}. 
Moreover, the derivation is generally applicable to a variety of three-body 
problems in arbitrary dimensions, in particular, the analytical expressions 
of the same form are obtained for three identical bosons~\cite{Kartavtsev99} 
and for three two-component fermions~\cite{Kartavtsev06} in 3D. 
All the considerations and the approach used are equally applicable to 
description of three identical 1D bosons, for which the exact solution is 
known. 
For these reasons, a brief discussion and numerical calculations for the 1D 
case are presented in the Appendix to make comparison with 2D results and 
to check the numerical procedure. 

The analytic expressions are used to analyze all the terms of HREs 
in the asymptotic region, thus obtaining the asymptotic form of the total 
wave function both for large and small inter-particle separation. 
In this respect, the universal dependence~(\ref{psiaszero}) is obtained for 
the total wave function in the vicinity of the triple-collision point with 
the leading term $\sim \ln^3\! \rho $ and the inter-particle correlations 
given by $\ln x_1 x_2 x_3 $. 
The large-$R$ asymptotic expansions are not uniform in channel number $n$, 
therefore, the explicit dependence on $n$ is deduced, which reveals 
convergence of the eigenpotentials and coupling terms to the limiting 
functions of $R - \ln{n} = \ln(\rho /n)$ at large $n$. 
The convergence is rather slow and the next-order term ($\sim n^{-1/2}$) in 
the large-$n$ expansion is periodic in $n$ with period 3; this is displayed 
by observing three families of eigenpotentials and coupling terms, namely, 
for different n mod 3. 
The asymptotic dependence on $n$ is used to study the effect of the channel 
coupling and to shed light on the convergence of the results with increasing 
number of HREs. 
One of the reasons for slow convergence is the long-range polarization tail 
$\sim e^{-2R} \sim \rho^{-4}$ of the first-channel effective potential and 
the same order long-range term which arises due to coupling with the upper 
channels. 
As a result, one needs to take into account a large number of HREs to study 
the long-range behaviour of the wave function and the next-to-leading terms of 
the low-energy effective-range expansion for $2 + 1$ scattering. 

The  universal constants, viz., the ground-state and excited-state three-body 
binding energies and the $2 + 1$ scattering length, are calculated 
with high precision by the numerical solution of HREs. 
The binding energies are in excellent agreement (within the declared accuracy) 
with those obtained in the momentum-space calculations~\cite{Hammer04}, which 
underlines equivalence of two essentially distinct models. 
The low-energy scattering of the dimer off the third particle is completely 
described by the precise $2 + 1$ scattering length. 

In summary, universal low-energy properties of three identical two-dimensional 
bosons are considered within the framework of the BCM used to describe 
two-body interactions. 
The approach used is based on the solution of a system of HREs, all the terms 
of which are derived in the analytical form. 
The derivation is quite general and can be applied to a number of problems, 
especially if the interaction is described within the framework of the BCM. 
The asymptotic form of the solutions of HREs is obtained, which allows one 
to describe the wave function both at large and small inter-particle 
separations. 
The binding energies and the $2 + 1$ scattering length of high precision are 
numerically calculated. 

\appendix
\section{Three one-dimensional particles}

In this appendix, the three-body problem in 1D is considered to demonstrate 
general applicability of the approach used, to check the numerical accuracy, 
and to compare convergence of the 1D and 2D calculations. 
The choice is based on well-known exact solubility of the one-dimensional 
N-body problem with the zero-range interactions~\cite{McGuire64,Lieb63}). 
As usual, the problem becomes parameterless by introducing the natural units 
$\hbar = m = 1$ and by choosing the potential strength to fix at unit values 
both the two-body binding energy, $\epsilon_2 = 1$, and the two-body 
scattering length, $a = 1$. 
The exact result for the binding energy of n identical particles in 
1D~\cite{McGuire64,Lieb63} is $\epsilon_{n} = \frac{1}{6}n(n^2 - 1)$. 
One should also mention that the ground-state wave function of three identical 
particles is of a simple form $\Psi_{gs} = C\exp(-\sum_k |x_k|)$, where 
the scaled Jacobi coordinates $x_i$ and $y_i$ are introduced similar to 
the above-discussed 2D case. 
The solution at the threshold energy $E = -1$ determines the wave function 
of three particles $\Psi_{sc} = \sum_k\exp(- |x_k|) - 
4\exp(-\frac{1}{2}\sum_k |x_k|)$, which entails infiniteness of the $2 + 1$ 
scattering length or existence of the zero-energy virtual 
state~\cite{Amaya-Tapia98}. 

Thereafter, the approach described in the paper is applied to calculate
the three-body binding energy $\epsilon_3$ and the $2 + 1$ scattering length
$A$ of three identical particles in 1D.
The wave function satisfies either the equation
\begin{equation}
\label{shred1dim}
\left[\frac{\partial^2}{\partial x^2} + \frac{\partial^2}{\partial y^2} +
2\sum_{i = 1}^3\delta(x_i) + E\right]\Psi = 0 \ ,
\end{equation}
where the zero-range interaction is a sum of the Dirac $\delta$-functions, 
or the free equation complemented by the boundary condition that
can be written for the each pair of the identical particles as
\begin{eqnarray}
\label{bound1dim1}
\lim_{x\to \pm 0} \left[ \frac{d}{d x} \pm 1\right]\Psi = 0\ .
\end{eqnarray}
Similar to Section~\ref{hrexpansion}, one introduces the variables $\rho$
and $\alpha_i$ and expands the wave function
\begin{equation}
\label{Psi1dim}
\Psi = \rho^{-1/2} \sum_{n=1}^{\infty}
f_n(\rho)\Phi_n(\alpha , \rho) \
\end{equation}
in a set of eigenfunctions $\Phi_n(\alpha , \rho)$ on a circle of constant 
$\rho $, which leads to the systems of ordinary differential equations 
for the functions $f_n(\rho)$ which are analogous to Eqs.~(\ref{system1}), 
(\ref{system2}).
Eigenpotentials in these systems are defined by the solution of the 
eigenvalue problem on a circle and the coupling terms are defined 
by the analytical expressions of the same form~(\ref{Qn3}), (\ref{Pn2}), 
and (\ref{Pn4}) as in 2D, provided the derivatives are taken over $\rho$. 
Recall that the derivation of the analytical expressions for the coupling 
terms in Section~\ref{Derivation} is equally applicable in 1D. 

For the symmetry reasons, the eigenvalue problem on a circle can be solved 
in the interval $0 \leq \alpha_i \leq \pi/6$ by imposing the zero boundary 
condition $\displaystyle\frac{\partial\Psi} {\partial \alpha} = 0$ at 
$\alpha = \pi/6$ and the boundary condition at $\alpha = 0$, 
\begin{eqnarray}
\label{bound1dim2}
\lim_{\alpha\to 0} \left[ \frac{d}{d \alpha} + \rho\right]\Psi = 0 \ 
\end{eqnarray}
which follows from Eq.~(\ref{bound1dim1}). 
The solutions of the eigenvalue problem on a circle satisfying the equation 
$\left(\displaystyle\frac{\partial^2}{\partial \alpha^2} + \xi_n^2 \right) 
\Phi_n(\alpha , \rho) = 0 $ take a simple form $\Phi_n(\alpha , \rho) = 
B_n\cos (\alpha - \pi/6)\xi_n $, where the eigenvalues $\xi_n(\rho)$ are 
defined by the transcendental equation, 
\begin{eqnarray}
\label{eigval1dim}
\xi + \rho\cot\frac{\pi}{6}\xi = 0 \ . 
\end{eqnarray}
Due to simple dependence $\xi(\rho)$~(\ref{eigval1dim}), one can derive 
simple analytical expressions for the coupling terms, for example, 
\begin{equation}
\label{Pnn1dim}
P_{nn} = \frac{\cos^4 \frac{x}{2} 
\left[x \left(x^2 - 3\right) \left(x-2\sin x \right) - 
6x^2 \cos x+ 3\sin^2 x\right]} {3x^2(x + \sin x)^4} \ , 
\end{equation}
where $x = \displaystyle\frac{\pi}{3}\xi_n(\rho)$, as follows from 
Eq.~(\ref{Pn4}). 

Similar to the 2D case, the numerical solution of the HREs 
in the form~(\ref{system1}),~(\ref{system2}) with zero boundary conditions 
gives the three-body binding energy, whereas the solution at the threshold 
energy $E = -1$ gives the $2 + 1$ scattering length. 
In the latter case the asymptotic form of the wave function is a product of 
the two-body bound-state wave function $\varphi_2(x) = \exp(-|x|)$ and 
the function $F = 1 - \frac{\sqrt{3}}{2}\frac{y}{A}$, which determines 
the asymptotic form of the first-channel function $f_1(\rho) = \rho^{-1/2} 
(1 - \frac{\sqrt{3}}{2}\frac{\rho}{A})$. 
As shown in Table~\ref{tab11dim}, the calculated binding energy rapidly 
converges to the exact value $\varepsilon_3 = 4$ with increasing $N$, whereas 
the calculated scattering length rapidly grows with $N$, which manifests 
infiniteness of the exact scattering length. 
\begin{table}[htb]
\caption{\label{tab11dim}
The three-body binding energy $\epsilon_3$ (in units of the two-body binding 
energy) and the $2 + 1$ scattering length $A$ for identical bosons in 1D. 
A number of HREs is denoted by $N$ and the superscripts U and L mark 
the results obtained by solving HREs of the form~(\ref{system1}) and 
(\ref{system2}), respectively. }
\begin{tabular}{c@{\hspace{2mm}}c@{\hspace{2mm}}c@{\hspace{2mm}}cc}
 $N$ & $\epsilon_3^U$ & $\epsilon_3^L$  & $ A $ \\
$ 1$ & $3.99934308$ & $4.00728928$ & $ 3.32000 \cdot 10^2 $ \\
$ 2$ & $3.99998993$ & $4.00055763$ & $ 7.9633 \cdot 10^3 $ \\
$ 3$ & $3.99999902$ & $4.00013463$ & $ 4.555 \cdot 10^4 $ \\
$ 5$ & $3.99999994$ & $4.00002429$ & $ 3.31 \cdot 10^5 $ \\
$ 7$ & $3.99999999$ & $4.00000816$ & $ 1.1 \cdot 10^6 $ \\
$ 9$ & $4.00000000$ & $4.00000367$ & $ 2.6 \cdot 10^6 $  \\
$12$ & $4.00000000$ & $4.00000149$ & $ 7.9 \cdot 10^6 $ \\
$15$ & $4.00000000$ & $4.00000074$ & $ > 10^7 $ \\
\end{tabular}
\end{table}
Both $\epsilon_3^U$ and $\epsilon_3^L$ are fairly well fitted to 
the $a + b/N^c$ dependence with the fitted values of power $c \approx 6$ and 
$c \approx 4$, respectively. 
The fitting of the scattering-length dependence on $N$ shows that 
the calculated $A$ grows as $N^3$. 
A better precision of the 1D calculation in comparison with the 2D one is 
basically due to a simple form~(\ref{bound1dim2}) of the eigenvalue equation 
that provides a better accuracy of the eigenpotentials. 
Both for 1D and 2D calculations, the second type of truncation of the HREs 
provides energies converging to the exact values from below. 

\bibliography{twodim3b}
\end{document}